\documentclass[a4paper,11pt]{article}
\pdfoutput=1 

\usepackage{jcappub} 

\usepackage[T1]{fontenc} 
\usepackage{subfig}

\def\be{\begin{equation}}
\def\ee{\end{equation}}
\def\barr{\begin{array}}
\def\earr{\end{array}}
\def\bea{\begin{eqnarray}}
\def\eea{\end{eqnarray}}
\def\bfig{\begin{figure}}
\def\efig{\end{figure}}

\def\eqn#1{eq.\ (\ref{#1})}
\usepackage{amsmath}
\newcommand{\fig}[1]{Fig.\ \ref{#1}}
\def\eqn#1{eq.\ (\ref{#1})}

\def\bea{\begin{eqnarray}}
\def\eea{\end{eqnarray}}

\def\be{\begin{equation}}
\def\ee{\end{equation}}
\def\barr{\begin{array}}
\def\earr{\end{array}}
\def\bfig{\begin{figure}}
\def\efig{\end{figure}}

\def\eqn#1{eq.\ (\ref{#1})}
\def\fig#1{fig.\ (\ref{#1})}

\newcommand{\ic}{{IceCube\,\,}}

\title{Flavour specific neutrino self-interaction: $H_0$ tension and IceCube}

\author[a]{Arindam Mazumdar,}
\author[b]{Subhendra Mohanty,}
\author[b,c]{Priyank Parashari}

\affiliation[a]{Centre for Theoretical Studies, Indian Institute of Technology, Kharagpur 721302, India}
\affiliation[b]{Physical Research Laboratory, Ahmedabad, 380009, India}
\affiliation[c]{Centre for High Energy Physics, Indian Institute of Science, C. V. Raman Avenue, Bengaluru 560012, India}
\emailAdd{arindam.mazumdar@iitkgp.ac.in}
\emailAdd{mohanty@prl.res.in}
\emailAdd{ppriyank@iisc.ac.in}

\abstract{
Self-interaction in the active neutrinos is studied in the literature to alleviate the $H_0$ tension.
Similar self-interaction can also explain the observed dips in the flux of the neutrinos coming from
the distant astro-physical sources in IceCube detectors.
In contrast to the flavour universal neutrino interaction considered for solving the $H_0$ tension, which is ruled out from particle physics 
experiments, we consider flavour specific neutrino interactions.
 We show that the values of self-interaction
coupling constant and mediator mass required for explaining the IceCube dips are inconsistent with the 
strong  neutrino self-interactions preferred by the combination of BAO, HST 
and Planck data. However, the required amount of self-interaction between tau neutrinos ($\nu_\tau$) in inverted hierarchy for explaining 
\ic dips is 
consistent with the moderate self-interaction region of cosmological bounds at 1-$\sigma$ level. For the case of other interactions and 
hierarchies, the \ic preferred amount of self-interaction is consistent with moderate self-interaction region of cosmological bounds at 
2-$\sigma$ level only. 
}


\begin{document}
\maketitle
\flushbottom

\section{Introduction}\label{intro}
Self-interaction in between the neutrinos has been an active topic of interest in different sectors of 
cosmology~\cite{Lancaster:2017ksf,Oldengott:2017fhy,Kreisch:2019yzn,Park:2019ibn,Barenboim:2019tux,Mazumdar:2019tbm, 
Blinov:2020hmc,He:2020zns}, 
astro-physics~\cite{Das:2017iuj,Ko:2020rjq} and laboratory based neutrino experiments~\cite{Blum:2014ewa,Blinov:2019gcj,Brdar:2020nbj}. In 
the recent years it has been studied extensively in the context of $H_0$ tension~\cite{Kreisch:2019yzn,Park:2019ibn}. Moreover, 
if 
this self-interaction is mediated by some MeV scale boson then the resonant on-shell production of mediator in the collision between the 
astro-physical neutrinos and the cosmic neutrino background can create some signature in the observed \ic PeV neutrino 
flux~\cite{Ng:2014pca,DiFranzo:2015qea,Shoemaker:2015qul,Bustamante:2020mep}. In this paper we check if these two types of applications of 
self-interaction in neutrinos are consistent with each other or not.       

There is a discrepancy between the determination of the Hubble constant $H_0$ from Planck \cite{Aghanim:2018eyx} (which 
assumes the $\Lambda$CDM cosmology) 
and those from local measurements based on distance ladder and time delay in lensing observations which points to new physics beyond the  
$\Lambda$CDM model \cite{Efstathiou:2013via,Bernal:2016gxb,Verde:2019ivm,DiValentino:2020zio}. The Planck observation finds the value 
$H_0 = (67.27 \pm 0.60)$ km/s/Mpc which is in 4.4$\sigma$  disagreement compared for example to the  SHOES collaboration 
\cite{Riess:2019cxk} determination of $H_0 = (74.03 \pm 1.42)$ 
km/s/Mpc, based on the observations by the Hubble Space Telescope of Cepheids in the Large Magellanic Cloud. One of the ways to alleviate  
the $H_0$ tension is to have a large self-interaction between neutrinos \cite{Kreisch:2019yzn}. Self-interaction delays the free-streaming 
of neutrinos and neutrinos cluster at smaller length scales. This is compensated by increasing $H_0$. In ref. \cite{Kreisch:2019yzn} the 
best fit value of $H_0$ which is closer to the distance ladder values was obtained by taking an effective neutrino self-interaction  ${\cal 
L}= G_{\rm eff} (\bar \nu \nu )(\bar \nu \nu )$  with $G_{\rm eff}$ having two preferred values where moderate self-interaction (MI) has 
$\log_{10}(G_{\rm eff} {\rm MeV^2})=-3.90^{+1.0}_{-0.93}$ and strong self-interaction (SI)  has  $\log_{10}(G_{\rm eff} {\rm 
MeV^2})=-1.35^{+0.12}_{-0.066}$. 
In a similar analysis in ref.~\cite{Park:2019ibn}, it is shown that when considering only WMAP data, which measures TT anisotropy spectrum 
up to multipole $l \leq 1200$, the bimodal peaks in the probability of $\log_{10}(G_{\rm eff} {\rm MeV^2})$ disappears and neutrino self 
interactions are consistent with zero. The bimodal distribution of $\log_{10}(G_{\rm eff} {\rm MeV^2})$ appears when Planck TT and TE data 
between $1200 \leq l \leq 2500$ are included. 
     
In ref.~\cite{Kreisch:2019yzn}  and  \cite{Park:2019ibn} (and in earlier studies of CMB with neutrino self-interactions 
\cite{Oldengott:2017fhy}, \cite{Lancaster:2017ksf}) the neutrino self-interaction that is considered is identical for all neutrino 
flavours. 
It has been pointed out  that such large flavour universal couplings of neutrinos which implies mediator masses of $\cal O( {\rm MeV})$ is 
severely constrained from particle physics. Strong bounds on $\nu$ self-interactions come from meson decays \cite{Blinov:2019gcj, 
Lyu:2020lps}, neutrino-less double beta decay \cite{Deppisch:2020sqh} and from $Z$ and $\tau$ decays \cite{Brdar:2020nbj}. Neutrino-less 
double beta decays rule out Majoron mediated $\nu_e$ interactions as a solution to the $H_0$ tension \cite{Deppisch:2020sqh}. Decays of 
$\pi^+ / K^+$ to $e^+ \nu_e$ and $\mu^+ \nu_\mu$ put strong constrains on $\nu_e$ and $\nu_\mu$ interactions while the constraints on 
$\nu_\tau$ couplings 
are determined only from $D_s^+ \rightarrow \tau^+ \nu_\tau$ which is not so well constrained \cite{Blinov:2019gcj, Lyu:2020lps}. The $Z$ 
and $\tau$ invisible decay width give the strongest bounds for heavy scalar mediators ( $m_\phi > 300 {\rm MeV}$) couplings to $\nu_\tau$ 
\cite{Brdar:2020nbj}.  To summarize, particle physics allows the self-interactions of $\nu_\tau$ in the MI region ($m_\phi\sim 10 -100 
{\rm MeV}$)   while universal flavour coupling interaction explored by refs. \cite{Kreisch:2019yzn, Park:2019ibn} in context of $H_0$ tension is ruled out as shown in ref.~\cite{Blinov:2019gcj}. 

In this paper, we consider flavour specific self-interactions between neutrinos and 
study their effect on CMB power spectrum. 
We consider four cases : 
$\nu_e , \nu_\mu , \nu_\tau $ and the flavour universal self-interactions.  From the CMB analysis we find that there is no discernible 
difference between the $\nu_\tau$ self-interactions allowed from particle physics constraints and the universal flavour interactions. 
We find, similar to the earlier papers~\cite{Oldengott:2017fhy,Lancaster:2017ksf,Kreisch:2019yzn,Park:2019ibn}, that $\log_{10}(G_{\rm eff} {\rm MeV^2})$ has a bimodal distribution in 
probability with a SI and MI peak. The allowed values of $N_{\rm eff}$ for the joint analysis of Planck CMB, BAO and HST data are 
significantly higher than three which ensures the larger best-fit values of the Hubble parameter.  The $H_0$ values obtained in the analysis of cosmological model having self-interacting neutrinos with Planck CMB, BAO and HST data reaches to the value of $H_0$ obtained from HST measurement. However, for this model, the $H_0$ inferred from the Planck CMB data alone does not change much from the $H_0$ values obtained in the $\Lambda$CDM model. 


Next, we test the flavour specific interactions with IceCube. It has been pointed out that PeV neutrinos from astrophysical sources can 
interact with cosmic neutrinos and produce an on shell mediator $\nu \nu \rightarrow \phi$, when the neutrino energy $E_\nu = m_\phi^2/ 
(2 m_\nu)$ and neutrino self-interactions with MeV mass mediators can have a signature in the  neutrino spectrum observed at IceCube \cite{ 
Ioka:2014kca, DiFranzo:2015qea,Chauhan:2018dkd, Kelly:2018tyg, Mohanty:2018cmq, Creque-Sarbinowski:2020qhz, Bustamante:2020mep}.  The 
resonant absorption of 
astrophysical neutrinos will show up as dips in the IceCube flux and this may explain the gap in the IceCube observations between $E_\nu = 
$400 TeV -1 PeV \footnote{Very recently, the \ic collaboration has reported the observation of Glashow-resonance event at $E_\nu =6.3$ PeV~\cite{IceCube:2021rpz}. When we were doing the analysis for this work. There was no observation of the Glashow resonance; that is why we have not used that in this work. However, we plan to include this in future analyses.
}. Since  the neutrino interactions are 
defined in the flavour basis while the resonance is in the mass basis, the neutrino mixing angles and mass hierarchies also play a crucial role 
in 
the IceCube spectrum. We find that all the flavour specific interactions show similar pattern in the \ic flux data, although these patterns 
are highly different from the universal interaction. We also find that, for the flavour specific and the universal self-interaction, the cosmological allowed SI 
region of interaction is ruled out by IceCube data while the MI region is partially consistent with IceCube for both the inverted and 
normal neutrino mass hierarchies.

The paper is organised as follows. In Section \ref{selfinteractions} we discuss the models of flavour specific self-interactions and 
discuss 
the scattering cross-section in the low energy limit relevant for CMB anisotropy, neutrino free-streaming and the high energy resonant 
interactions which are relevant for IceCube  neutrinos.  In Section \ref{CMBconstraints} we discuss the constraints on effective 
interaction strength ($G_{\rm eff}$) for the flavour specific and universal interactions from Planck CMB, BAO and HST data. 
 In Section \ref{Neutrinopropagation} we discuss the propagation of high energy neutrinos in cosmic neutrino background with resonant 
scattering. 
In Section \ref{IceCube} we apply the analysis of high energy neutrino propagation to IceCube flux considering the four flavour specific 
neutrino interactions for the two mass hierarchies. In Section \ref{Conclusions}, we conclude the viability of neutrino 
interaction models in the light of cosmological and IceCube data.

%
\section{Self-interaction in neutrinos }\label{selfinteractions}

Neutrino self-interactions can be mediated by scalars and gauge bosons which are motivated from different particle physics 
models\cite{Chikashige:1980ui, Gelmini:1980re, Georgi:1981pg,Gelmini:1982rr,Nussinov:1982wu,Dey:2020fbx}. The 
lepton number is conserved in the standard model and in the extensions of the standard model where lepton number is broken spontaneously, there 
are scalars called Majorons which arise from Goldstone bosons of the lepton number symmetry breaking \cite{Arias-Aragon:2020qip}. 
Neutrino self-interactions can also arise from the gauged lepton number and anomaly free extensions of the standard model where light gauge 
bosons couple to the neutrinos and evade all other experimental constraints~\cite{Berbig:2020wve}. To analyse the low energy particle 
physics and CMB constraints, it is enough to work in the effective theory framework 
\cite{Lyu:2020lps}. However, to analyse high energy IceCube interactions, where the resonance behaviour of cross-section is needed,  the 
full theory is required.  

The neutrino self-interactions are defined in the flavour basis ($\nu_\alpha$) but the high energy propagation and the neutrino 
free-streaming is analysed in the neutrino mass basis ($\nu_i$). Here Greek letters are for three different flavours $ e,\mu$ and $\tau$  
and 
Latin letters are for mass eigenstates which run from 1 to 3. 
To relate the couplings in the flavour basis to those in the mass basis, the PMNS mixing matrix defined as 
\begin{eqnarray}
 |\nu_\alpha\rangle = U_{\alpha i}|\nu_i\rangle\, ,
\end{eqnarray}
where the values of the components of $U_{\alpha i}$ have been taken from latest NuFit data~\cite{Esteban:2020cvm}.
Global analysis of the latest neutrino oscillation data provides us the values 
for all the oscillation parameters like mass squared difference $\Delta m_{ij}^2 = m_i^2 - m_j^2\,$ and the mixing angles. However, the 
existing data failed to give the correct sign of $\Delta m_{31}^2 $ or $\Delta m_{32}^2 $. Therefore, we have two mass hierarchies, namely 
normal hierarchy (NH) and inverted hierarchy (IH). In case of NH $m_1 < m_2 < m_3$ and in case of IH $m_3 < m_1 < m_2$. For both the 
hierarchies, we will assume the lowest neutrino mass to be $m_0$.

Self-interaction in between the active neutrino species can occur from gauge-interaction or Yukawa like interactions mediated by a 
scalar($\phi$) particle.  In case of Yukawa like interactions the Lagrangian can be written as
\begin{eqnarray}
 -\mathcal{L} = g_\phi \sum_{\alpha,\beta} g_{\alpha\beta} \phi \bar{\nu}_\alpha \nu_\beta \, ,
\end{eqnarray}
where $g_\phi$ is the coupling strength. In the mass basis this can be written as
\begin{eqnarray}
 -\mathcal{L} = g_\phi\sum_{i,j} g_{ij} \phi \bar{\nu}_i \nu_j \, ,
\end{eqnarray}
where $ g_{ij} = g_{\alpha\beta} U^{*}_{\alpha i} U_{\beta j}$.
Similarly for gauge-interactions the Lagrangian can be written as 
\begin{equation}
\label{sl}
- \mathcal{L} = g_X \sum_{\alpha,\beta}\bar{\nu}_{\alpha} g_{\alpha\beta}\gamma^\mu P_L \nu_{\beta} X_\mu\, ,
\end{equation}
where $g_X$ is the coupling strength. In terms of mass eigenstates this kind of interaction term becomes 
\begin{equation}
- \mathcal{L} = g_X \sum_{i,j} g_{ij} \bar{\nu}_i \gamma^\mu P_L \nu_j X_\mu\, .
\end{equation}
The matrix $g_{\alpha\beta}$ defines the flavour dependence of the interactions.
As discussed earlier, in this paper, we will work with four different types of flavour dependencies. Therefore $g_{\alpha\beta}$ will be 
$\delta_{\alpha\beta}$ for universal interaction and for interaction in a particular flavour it will be a diagonal matrix with only one 
among 
$g_{ee},g_{\mu\mu}$ or $g_{\tau\tau}$ set to be one. 

For both the scalar and vector exchange cases, for momentum-transferred smaller than the mediator mass, the neutrino self-interactions  can 
be described by the four-Fermi term
\begin{eqnarray}
 {\cal L} = {G}_{\rm eff}\, g_{\alpha\beta}g_{\gamma\delta}{\bar\nu_\alpha} \nu_\beta {\bar\nu_\gamma} \nu_\delta \, ,
\end{eqnarray}
where ${G}_{\rm eff}= g_\phi^2/m_\phi^2$ or ${G}_{\rm eff}= g_X^2/m_X^2$ for scalar and gauge boson exchange respectively. In the mass basis, the above Lagrangian takes the following form:
\begin{eqnarray}
 {\cal L} = G_{\rm eff}\, g_{ij}g_{kl}{\bar\nu_i} \nu_j {\bar\nu_k} \nu_l \, .
\end{eqnarray}
Therefore, for a process like $ \bar{\nu}_i \nu_j \rightarrow \bar{\nu}_k \nu_l $, the cross-section will be given as $\sigma_{ijkl} = |g_{kl}|^2|g_{ij}|^2 {G}_{\rm eff}^2$. After summing over the final states, the cross-section is given as
\begin{eqnarray}\label{eqn:sig2_eff}
\sigma_{ij}= \sum_{k,l} |g_{kl}|^2|g_{ij}|^2 {G}_{\rm eff}^2 = \mathcal{A} |g_{ij}|^2 {G}_{\rm eff}^2\,,
\end{eqnarray}
where $\mathcal{A}=\sum_{k,l} |g_{kl}|^2$. For a diagonal $g_{\alpha \beta}$ considered in this work, $\mathcal{A}=\sum_{k,l} U_{k \alpha}^\dagger g_{\alpha\beta} U_{\beta l}$.
%
For the CMB analysis where the momentum transfer in neutrino scatterings are smaller than ${\rm MeV}^2$, the Four-Fermi effective operator is adequate for the 
analysis. However, when we consider interactions of high-energy astrophysical neutrinos with the cosmic neutrino background, the effective 
Four-Fermi interaction is not applicable and one must do the calculations for the full theory with the mediator mass playing a crucial role 
specially  near resonant scattering energies.  
The high energy neutrino-neutrino scattering cross-section due to scalar exchange is given 
by~\cite{Goldberg:2005yw,Farzan:2014gza,Creque-Sarbinowski:2020qhz}
\begin{equation}\label{eqn:sig4}
\sigma_{ijkl} = \sigma \left( \bar{\nu}_i \nu_j \rightarrow \bar{\nu}_k \nu_l \right) =
\frac{1}{4 \pi} |g_{kl}|^2|g_{ij}|^2  \frac{g_\phi^4 s_j}{(s_j - m_{\phi}^2)^2 + m_{\phi}^2 \Gamma_{\phi}^2}\, ,
\end{equation}
where $s_i=2 E_i m_i$ and $\Gamma_\phi= g_\phi^2\sum_{i,j}|g_{ij}|^2 m_\phi/4\pi$ is the decay width of the mediator.   For vector 
interactions the cross-section has the same Breit-Wigner form with different constants in the prefactor.  When summed over 
the final states this turns out to be 
\begin{eqnarray}\label{eqn:sig2}
\sigma_{ij}= \sigma \left( \bar{\nu}_i \nu_j \rightarrow \bar{\nu} \nu \right) = \sum_{k,l}\sigma_{ijkl}\, .
\end{eqnarray}
Therefore, we can see that for those energies where $s_i$ becomes equal to the $m_\phi^2$, the scattering cross-section becomes maximum. 
These energies are called resonant energies and denoted by ${E_R}_i=m_\phi^2/2m_i$. In section~\ref{Neutrinopropagation} we will show 
that corresponding to 
these resonant energies the absorption rates of astrophysical neutrinos reach the maximum value and we find the dips in the neutrino spectrum in 
those energy values. 

\section{Constraints from cosmological data set}\label{CMBconstraints}
%
We now turn our focus on  the cosmological perturbation theory in the presence of self-interaction in massive neutrinos. Massive 
neutrinos play an important role in the evolution of cosmological perturbations as well as in the evolution of background cosmology.
Generally, neutrinos free stream in the baryon-photon fluid of early universe and their free streaming length mainly depends on their 
masses. However, if the neutrinos have self-interaction in between them then that reduces their free streaming length and makes the 
neutrinos clump together more and more than the free neutrinos. The mass of the neutrinos on the other hand not only modifies the free 
streaming length but also modifies the Hubble parameter. That is why the effects of self-interacting neutrinos and the massive 
neutrinos on CMB are not same. The neutrino interactions are defined in the flavour basis while the neutrino density perturbations are analysed in 
the 
mass basis. To study this, we need to calculate the density matrix for the neutrinos which can be written as 
$\rho_{\alpha\beta}=|\nu_\alpha\rangle\langle\nu_\beta|$ and this transforms to mass basis as    
\begin{eqnarray}
 \rho_{\alpha\beta} = U_{\alpha i} \rho_{ij} U_{j \beta}^* ~.
\end{eqnarray}
Next, we need to find out the Boltzmann hierarchy equations for the massive neutrinos in the presence of a collision term arising due to the self-interacting neutrinos, which can be written as~\cite{Ma:1995ey}
\begin{equation}\label{eq:bol_col}
\frac{\partial \Psi_i}{\partial \tau} + i \frac{q(\vec{k} \cdot \hat{n})}{\epsilon} \Psi_i + \frac{d \ln f_0}{d \ln q} \left[ \dot{\eta} -
\frac{\dot{h}+6\dot{\eta}}{2} \left(\hat{k} \cdot \hat{n} \right)^2 \right] = \frac{1}{\bar{f_0^i}}\frac{\partial f_i}{\partial \tau}
\end{equation}
where $\bar{f_0^i}=\sum_\alpha f_0|U_{\alpha i}|^2\rho_{\alpha\alpha}$ with $f_0$ being the zeroth order Fermi-Dirac distribution function~\cite{Mazumdar:2019tbm}, and $\Psi$ is the scalar perturbation in the distribution function. If there is no self-interaction, the collision term on the right hand side of this equation will be set to zero. The scalar perturbation $\Psi$ in the distribution function is expanded in terms of the Legendre polynomials as,
\begin{equation}\label{expansion}
\Psi(\vec k,\hat n, q,\tau) = \sum_{\ell =0}^{\infty} (-i)^\ell (2\ell +1)\,\Psi_\ell (\vec k, q,\tau) P_\ell(\hat k . \hat n)\, . 
\end{equation}
Such expansion of $\Psi$ leads to the Boltzmann hierarchy equations consisting of the individual equations corresponding to each multipole.


The collision term for the Boltzmann hierarchy equations in the case of neutrino has been considered in literature in many different ways. Refs.~\cite{Oldengott:2014qra,Oldengott:2017fhy,Kreisch:2019yzn} provide the exact calculation, refs.~\cite{Archidiacono:2011gq,Smith:2011es} provide the effective fluid equation and refs.~\cite{Lancaster:2017ksf,Song:2018zyl,Mazumdar:2019tbm} obtain the collision term under the relaxation time approximation. We follow the relaxation time approximation since it is computationally less time taking. According to relaxation time approximation the collision term with $\ell\ge 2$ is given as
\begin{eqnarray}\label{eq:app_rela}
\frac{1}{\bar{f_0^i}}\frac{\partial f_i}{\partial \tau}=-\Gamma_{ij}\Psi_j\,.
\end{eqnarray}
Here  $\Gamma_{ij}$ is the scattering rate between neutrinos, which is defined as
\begin{equation}
\Gamma_{ij} = {a}n_\nu \langle\sigma_{ij} v\rangle
\end{equation}
where $\sigma_{ij} = \mathcal{A}|g_{ij}|^2 G_{\rm eff}^2 T_\nu^2$ is the effective cross-section for the self-interacting neutrinos on the CMB relevant scales (see \eqn{eqn:sig2_eff}). For a diagonal matrix $g_{\alpha \beta}$ as considered in this paper, $|g_{ij}|^2 = U_{i \alpha}^\dagger g_{\alpha\beta} U_{\beta j}$. Therefore, the scattering rate can be written as
\begin{equation}
\Gamma_{ij} = U_{i \alpha}^\dagger g_{\alpha\beta} U_{\beta j} \frac{3}{2}\frac{\zeta(3)}{\pi^2}{a}G_{\rm eff}^2 {T_\nu^5 \rho_{kk}}\mathcal{A}\,,
\end{equation}
where we have assumed the neutrino temperature ($T_\nu$) to remain the same for the three neutrino mass eigenstates in the presence of self-interaction. Here, the diagonal elements of the neutrino density matrix $\rho_{kk}$ correspond to the neutrino number density $n_\nu$ normalized to the number density $n_\nu^*$ in a reference model, which is assumed to be the $\Lambda$CDM model plus three active neutrinos without self-interaction~\cite{Song:2018zyl}. As a result, a factor of $\rho_{kk} $ arises in the scattering rate. However, this factor is unimportant here as it is equal to one in this case. But, as shown in ref.~\cite{Song:2018zyl,Mazumdar:2019tbm}, the value of this factor deviates from one in the case of self-interacting light sterile neutrinos. Let us now rewrite the scattering rate in the relaxing time approximation as~\cite{Hannestad:2000gt}  
\begin{equation}\label{eq:gam_f}
\Gamma_{ij} = U_{i \alpha}^\dagger g_{\alpha\beta} U_{\beta j} \tau_\nu^{-1} ~~~~ {\rm with} ~~~~~ \tau_\nu^{-1}=\frac{3}{2}\frac{\zeta(3)}{\pi^2}{a}G_{\rm eff}^2 {T_\nu^5 \rho_{kk}} \mathcal{A}\,,
\end{equation}

However, the exact form of collision term for $\ell = 0$ and $\ell = 1$ in relaxation time approximation for self-interaction in different species are not available in literature. Therefore we move forward here with the following assumptions. 
According to ref.~\cite{Oldengott:2014qra}, the exact collision term in $\ell = 0$ and $\ell = 1$ equations 
depends on the differences between the distribution functions of different mass eigenstates. Let us investigate if that term would have any significant contribution. The $G_{\rm eff}$ values allowed from CMB data has been reported to be less than $\sim 10^{-1}$ MeV$^{-2}$~\cite{Kreisch:2019yzn}. For this level of interaction  
we find that the self-interaction rate of the neutrinos becomes smaller than the Hubble expansion rate at a temperature $T\sim20$ eV. However, neutrino masses that we will consider in this paper will be limited to less than 1 eV.  Therefore, when the self interaction plays its role in Boltzmann equations, the neutrinos will be completely in the relativistic regime and they will have the same distribution function for all the mass eigen states. Moreover, we have already considered that the neutrinos are in thermal equilibrium with each other which ensures that number of each species should be conserved. Therefore we set collision term to zero in $\ell = 0$ case. However, there can be a velocity slip kind of term in the equation for $\ell =1$ for conservation of overall momentum. Such an term will introduce a term like $\Gamma_{ij}(\Psi_{j,1} - \Psi_{i,1})$ in the equation for $\ell =1$ for our case. We checked the effect of such a term on the CMB power spectrum and found it to be indistinguishable from the power spectrum where this term is not considered. Therefore, for doing the MCMC analysis, we set this term to zero. In a very recent study similar assumption is taken by setting the neutrino masses to zero~\cite{Das:2020xke}.

Using Eqs. \eqref{eq:gam_f} and \eqref{eq:app_rela} in \eqn{eq:bol_col} and \eqn{expansion}
the Boltzmann hierarchy equations takes the following form~\cite{Forastieri:2017oma,Song:2018zyl}  
\begin{subequations}\label{eq:scalar-pert}
\begin{align}
&\dot{\Psi}_{i,0} = -\frac{q k}{\epsilon} \Psi_{i,1} + \frac{1}{6} \dot{h} \frac{d \ln{f_0}}{d \ln{q}} \, ,\\
&\dot{\Psi}_{i,1} = \frac{qk}{3 \epsilon} \left(\Psi_{i,0} - 2 \Psi_{i,2} \right) \, ,\\ 
&\dot{\Psi}_{i,2} = \frac{qk}{5 \epsilon} \left( 2 \Psi_{i,1}- 3 \Psi_{i,3} \right) - \left( \frac{1}{15} \dot{h} + \frac{2}{5}\dot{\eta} 
\right) \frac{d \ln{f_0}}{d \ln{q}} - \Gamma_{ij} \Psi_{j,2}\, ,\\
&\dot{\Psi}_{i,\ell} = \frac{qk}{(2\ell+1)\epsilon} \Big[ \ell \Psi_{i,(\ell-1)} - (\ell+1) \Psi_{i,(\ell+1)} \Big]  - \Gamma_{ij} 
\Psi_{j,\ell} \quad (\ell\ge 3)\, .
\end{align}
\label{eq:boltz_mass_int}
\end{subequations}
\begin{figure}[t]\centering
 \includegraphics[width=0.49\linewidth]{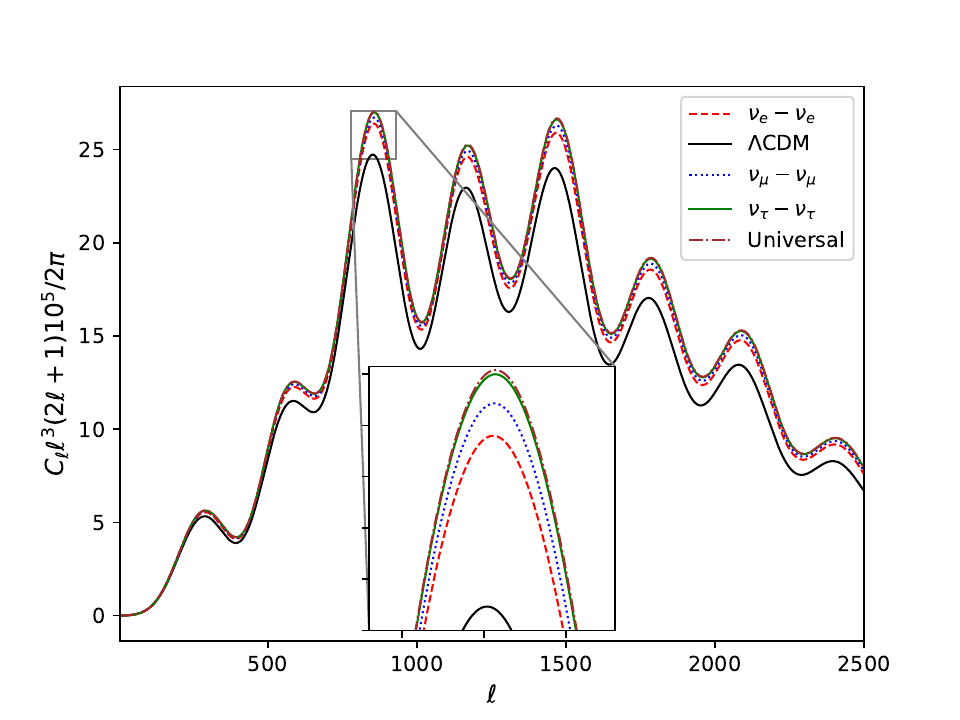}
 \includegraphics[width=0.49\linewidth]{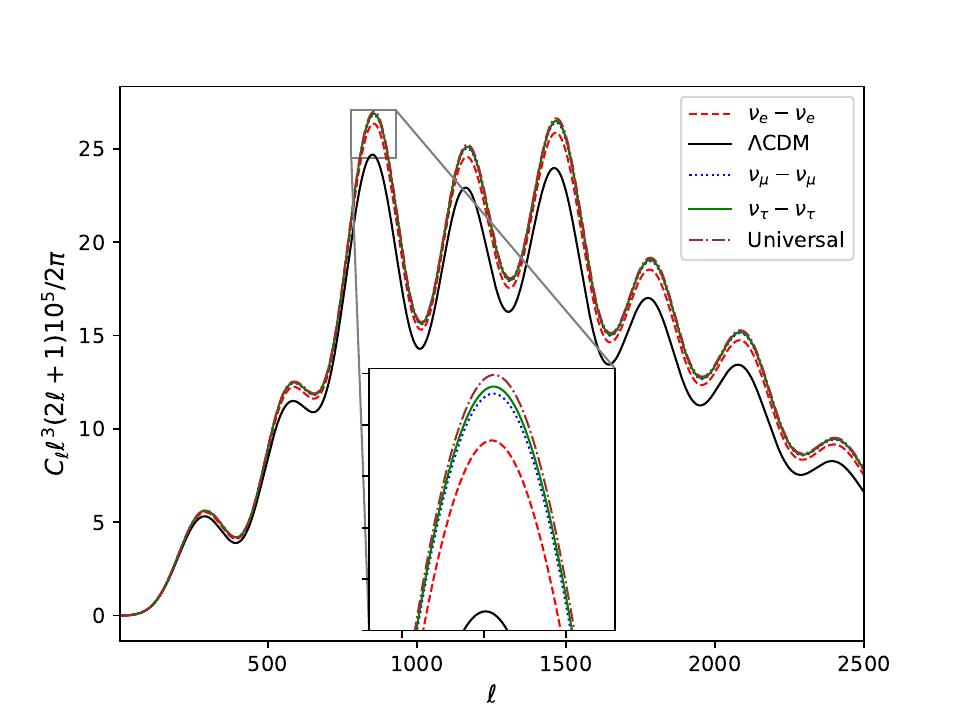}\\
 (a) Normal hierarchy \hspace*{0.33\linewidth}(b) Inverted hierarchy
 \caption{Effects of flavour specific interactions on CMB temperature power spectrum are shown. Power spectrum corresponding to universal 
interaction  almost overlaps with that of $\nu_\tau$-$\nu_\tau$ interaction for both the hierarchies. The value of $G_{\rm 
eff}$ has been fixed to $10^{-1.5}{\rm MeV}^{-2}$ for all these cases.}\label{cl_norm}
\end{figure}

We modified these equations (\eqn{eq:boltz_mass_int}) accordingly in the Boltzmann code CLASS~\cite{Lesgourgues:2011re} and have shown the 
effects of these interactions 
on CMB in \fig{cl_norm}. In general, self-interaction between neutrinos helps the small scale perturbations to grow therefore the height 
of the peaks in CMB power spectrum increases. However, the effect of flavour specific interactions on CMB 
spectrum shows that there are some minor differences between the effects of different interactions. These effects can be understood from 
the equations~\ref{eq:boltz_mass_int}. In the case of universal interaction when 
$g_{\alpha\beta}$ is equal to $\delta_{\alpha\beta}$ the $\Gamma_{ij}$ becomes a diagonal matrix. That means the growth of the scalar 
perturbation multipoles ($\Psi_\ell$) of one mass eigenstate depends only on that mass eigenstate. However, for the case of flavour 
specific interactions growth of $\Psi_\ell$ depends on other mass eigenstates too. Ultimately, the amount of the effect of 
self-interaction is determined by the quantity $\sum_{i,j}\Gamma_{ij}$. In case of universal interaction it is $3\tau_\nu^{-1}$. For 
$\nu_e$-$\nu_e$ 
interaction this becomes $2.308\,\tau_\nu^{-1}$, for $\nu_\mu$-$\nu_\mu$ $2.643\,\tau_\nu^{-1}$ and for $\nu_\tau$-$\nu_\tau$ it turns out 
to be 
$2.965\,\tau_\nu^{-1}$. In case of inverted hierarchy these numbers are $2.309\,\tau_\nu^{-1}$ for $\nu_e$-$\nu_e$, $2.809\,\tau_\nu^{-1}$ 
for 
$\nu_\mu$-$\nu_\mu$ and $2.88116\,\tau_\nu^{-1}$ for $\nu_\tau$-$\nu_\tau$. Therefore, we see that the effects of universal interaction and 
$\nu_\tau$-$\nu_\tau$ 
interaction on CMB are almost indistinguishable for both the hierarchies in \fig{cl_norm}.

\begin{figure}[h]
\centering{
\includegraphics[width=0.4\linewidth]{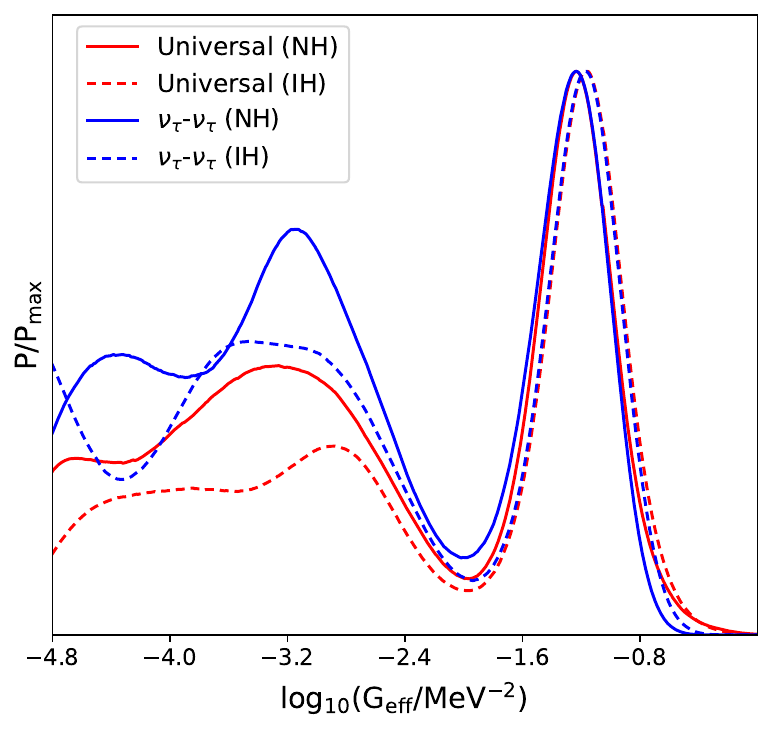}
 \includegraphics[width=0.4\linewidth,height=0.365\linewidth]{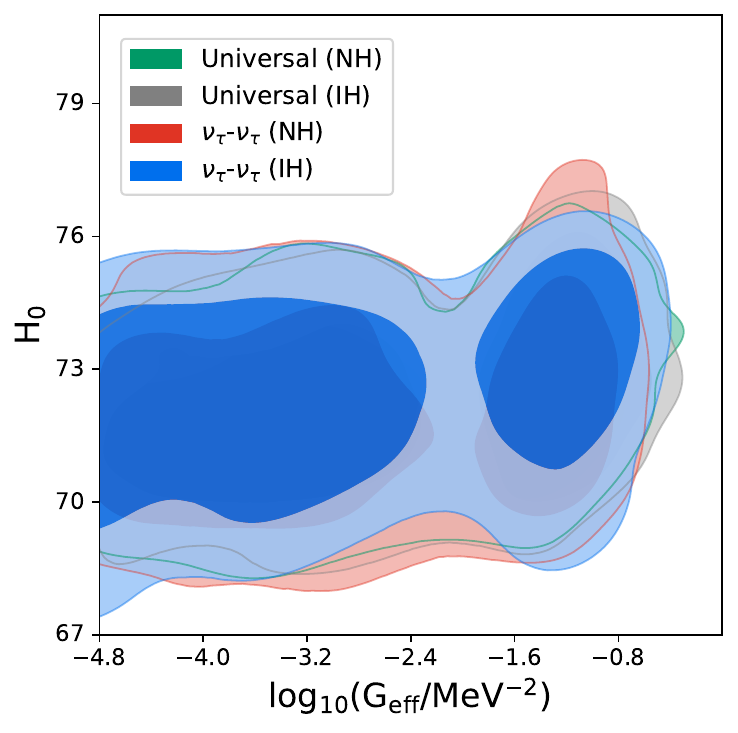}\\
(a) \hspace*{0.35\linewidth} (b) 
} 
 \caption{ (a) Posterior distribution of $G_{\rm eff}$ is bimodal in nature. (b) Inclusion of self-interaction leads to higher $H_0$ as compared to that in $\Lambda$CDM model with $N_{\rm eff}$ greater than three when Planck, BAO and HST data analysed jointly. There 
is no significant effect of hierarchies  or the flavour specific nature of the self-interaction on constraining 
parameters.}\label{cmb_posterior}
\end{figure}

 We proceed to constrain the parameter space of $G_{\rm eff}$ with Markov Chain Monte Carlo (MCMC) technique using 
MontePython~\cite{Audren:2012wb}.  We have used the 
Planck high-$\ell$ and low-$\ell$ likelihood, following ref~\cite{Aghanim:2018eyx,Akrami:2018odb} where high-$\ell$ consists of only $TT$ 
spectrum and $TT$,$TE$ and $EE$ spectrum is incorporated in low-$\ell$ likelihood\footnote{ In the literature, the high-$\ell$ polarization data have also been used to constrain neutrino self-interaction. However, we have only used high and low-$\ell$ data for temperature anisotropy and low-$\ell$ polarization data to be consistent with our previous work~\cite{Mazumdar:2019tbm} and to avoid extra computational time. The constrained value of $G_{\rm eff}$ obtained in our analysis is similar to the value found in other analyses where high-$\ell$ polarization is used. In this work, our main focus is to compare the $G_{\rm eff}$ values allowed from the cosmological observations with the best-fit values of the IceCube observation. Inclusion of high-$\ell$ polarization data may change the allowed $G_{\rm eff}$ values slightly, however, the conclusion of this work will not be affected by it.}. 
We have used two other data sets. One is baryon acoustic oscillation scale set by BAO-BOSS data of DR12 release~\cite{Alam:2016hwk} and 
another one is the measured value of $H_0$ by Hubble space telescope from the observations of Cepheid~\cite{Riess:2019cxk}. We will refer 
this combined data set as ``Planck+BAO+HST''.  In this analysis we have varied $N_{\rm eff}$, lowest neutrino mass 
$m_0$ and six standard cosmological parameters. The details of the MCMC analysis is provided in the appendix~\ref{app:numerical}.  

\begin{table}[h]
\begin{center}
\resizebox{1.0\textwidth}{!}{
\begin{tabular} {  l  c  c   c  c }
\hline
\hline
 Parameter & Universal Interaction (NH) & Universal Interaction (IH) & $\nu_\tau$-$\nu_\tau$ Interaction (NH) & $\nu_\tau$-$\nu_\tau$ 
Interaction (IH)\\
\hline 
{\boldmath$10^{2}\omega_{b }$} & $2.269^{+0.037}_{-0.033}         $ & $2.268\pm 0.036            $& $2.261\pm 0.037            $ & 
$2.267\pm 0.034            $\\

{\boldmath$\omega_{cdm }  $} & $0.1286^{+0.0058}_{-0.0068}$ & $0.1286\pm 0.0059          $& $0.1279^{+0.0051}_{-0.0060}$ & 
$0.1281^{+0.0054}_{-0.0065}$\\

{\boldmath$100\theta_{s } $} & $1.0412^{+0.0009}_{-0.0011}         $  & $1.0411^{+0.0010}_{-0.0013}$ & $1.0413^{+0.0010}_{-0.0011}$  & 
$1.0411\pm 0.0010          $\\

{\boldmath$\ln(10^{10}A_{s })$} & $3.062\pm 0.038            $ & $3.062\pm 0.037 $& $3.058\pm 0.037            $ & $3.056\pm 
0.035            $\\

{\boldmath$n_{s }         $} & $0.987\pm0.014   $ & $0.988^{+0.016}_{-0.013}   $& $0.984^{+0.016}_{-0.013}   $ & $0.987 \pm {0.014}  
$\\

{\boldmath$\tau_{reio }   $} & $0.057\pm0.016   $ & $0.058\pm 0.016            $& $0.056\pm 0.016            $ & $0.055\pm 0.016            
$\\

{\boldmath${m_0 } $} & $0.058^{+0.022}_{-0.051}   $ &  $0.059^{+0.024}_{-0.056}   $
& $0.052^{+0.018}_{-0.049}   $ & $0.053^{+0.019}_{-0.044}   $\\

{\boldmath$\log_{10}{G_{\rm eff} }      $} & $-3.48^{+0.94}_{-0.65}     $ & $-3.47^{+0.79}_{-0.75}      $ & $-3.44^{+1.0}_{-0.63}             $ & 
$-3.56^{+1.1}_{-0.79}   $\\

{\boldmath$N_{\rm eff}        $} & $3.76^{+0.31}_{-0.38}      $ & $3.78^{+0.29}_{-0.34}      $& $3.69^{+0.29}_{-0.33}      $ & 
$3.75 \pm {0.33}    $\\

{\boldmath$H_0             $} & $72.0^{+1.7}_{-1.9}               $  & $71.9\pm 1.7               $ & $71.6\pm 1.7               $ & 
$71.9^{+2.0}_{-1.8} $\\
\hline
\end{tabular}
}
\end{center}
\caption{1-$\sigma$ allowed values of all the parameters for the moderate self-interaction ({\bf MI}) for both universal and 
$\nu_\tau$-$\nu_\tau$ interactions.}\label{tab:mi_values}
\end{table}

\begin{table}[h]
\begin{center}
\resizebox{1.0\textwidth}{!}{
\begin{tabular} {  l  c  c   c  c }
\hline
\hline
 Parameter & Universal Interaction (NH) & Universal Interaction (IH) & $\nu_\tau$-$\nu_\tau$ Interaction (NH) & $\nu_\tau$-$\nu_\tau$ 
Interaction (IH)\\
\hline 
{\boldmath$10^{2}\omega_{b }$} & $2.237\pm 0.036         $ & $2.249\pm 0.035          $& $2.246\pm 0.041          $ & $2.250\pm 
0.035          $\\

{\boldmath$\omega_{cdm }  $} & $0.1344\pm0.0064$ & $0.1349\pm0.0066$& $0.1369^{+0.0073}_{-0.0063}$ & $0.1371^{+0.0078}_{-0.0047}$\\

{\boldmath$100\theta_{s } $} & $1.0457\pm 0.0012          $  & $1.0455^{+0.0011}_{-0.0014}          $ & $1.0453\pm 0.0011          $ & 
$1.0451^{+0.0009}_{-0.0010}          $\\

{\boldmath$\ln(10^{10}A_{s })$} & $3.002\pm 0.039         $ & $3.003^{+0.031}_{-0.037}            $& $3.002\pm 0.036            $ & 
$3.012\pm 0.035            $\\

{\boldmath$n_{s }         $} & $0.958\pm0.014   $ & $0.959\pm0.013   $& $0.961\pm 0.013 $ & $0.964\pm0.011   $\\

{\boldmath$\tau_{reio }   $} & $0.054\pm 0.016   $ & $0.055^{+0.013}_{-0.015}   $ & $0.049 \pm 16  $ & $0.054\pm 0.015  
  $\\

{\boldmath$m_0 $} & $0.080^{+0.039}_{-0.068}   $ & $0.083^{+0.045}_{-0.053}   $
& $0.083^{+0.043}_{-0.061}   $ & $0.091^{+0.062}_{-0.063} $\\

{\boldmath$\log_{10}{G_{\rm eff} }      $} & $-1.25^{+0.21}_{-0.18}     $ & $-1.17\pm 0.23    $& $-1.28^{+0.26}_{-0.17}     $& 
$-1.18^{+0.17}_{-0.12}     $\\

{\boldmath$N_{ \rm eff }        $} & $3.93\pm0.35      $ & $3.98\pm 0.36      $ & $4.06 \pm 0.35 $ & $4.10^{+0.39}_{-0.19}      
$\\

{\boldmath$H_0             $} & $72.7\pm 1.9               $  & $72.9\pm 1.7               $ & $73.3\pm 1.8               $ & 
$73.2\pm 1.8             $\\
\hline
\end{tabular}
}
\end{center}
\caption{1-$\sigma$ allowed values of all the parameters for the strong self-interaction ({\bf SI}) for both universal and 
$\nu_\tau$-$\nu_\tau$ interactions.}\label{tab:si_values}
\end{table}

As shown in the earlier literature~\cite{Lancaster:2017ksf,Oldengott:2017fhy,Kreisch:2019yzn} we also find that the Planck data along with 
BAO and HST data prefer a 
small region of strong interaction between the neutrinos in \fig{cmb_posterior}-(a). The posterior of $G_{\rm eff}$ is bi-modal. 
For quantifying the two regions of self-interaction we separate out the points from the posterior distribution which have $\log_{10}(G_{\rm 
eff})$ values greater than -1.95 and less than that. We use GetDist~\cite{Lewis:2019xzd} to extract the statistics of these peaks. The 
strong interaction region (we call SI onwards) has the value of $\log_{10}(G_{\rm eff}/{\rm MeV}^{-2})$ in one sigma range as
 $-1.25^{+0.21}_{-0.18}  $  for universal interaction in the normal hierarchy, $-1.17\pm 0.23$ for universal interaction in the inverted 
hierarchy, $-1.28^{+0.26}_{-0.17}$ for $\nu_\tau$-$\nu_\tau$ interaction in the normal hierarchy and $-1.18^{+0.17}_{-0.12}$ for 
$\nu_\tau$-$\nu_\tau$ interaction in the inverted hierarchy. For these specific interactions the bestfit values for $\log_{10}(G_{\rm 
eff}/{\rm MeV}^{-2})$ in the mildly interacting region (we call MI onwards) are $-3.48^{+0.94}_{-0.65}$, $-3.47^{+0.79}_{-0.75}  $, 
$-3.44^{+1.0}_{-063}$ and $-3.56^{+1.1}_{-0.79}$ respectively. The combined ``Planck+BAO+HST'' analysis in the case of varying $N_{\rm 
eff}$ can push the value of $H_0$ up to $73.3 \pm 1.8$ in the case of $\nu_\tau$-$\nu_\tau$ interaction in normal hierarchy. Similarly for other interactions also the best-fit value of $H_0$ becomes higher (see \fig{cmb_posterior}-(b)). 
 Values of all other parameters are presented in table~\ref{tab:mi_values} and table~\ref{tab:si_values}.

In \fig{cl_norm} we see that difference in between different types of interactions is almost indistinguishable unless magnified. Even after 
magnification there is almost no 
difference in between the effects of $\nu_\tau$-$\nu_\tau$ and universal interaction on CMB spectra. Similarly the MCMC 
bound on $G_{\rm eff}$ for the universal interaction are almost same to the bound in  $\nu_\tau$-$\nu_\tau$ interaction for both the 
hierarchies. With these bounds on $G_{\rm eff}$ and $m_0$ in hand  
we move forward to calculate the effect of neutrino self-interactions on \ic flux. 

\section{Neutrino absorption by Cosmic Neutrino Background}\label{Neutrinopropagation}
\begin{figure}
 \centering
 \includegraphics[width=0.24\linewidth,height=0.2\linewidth]{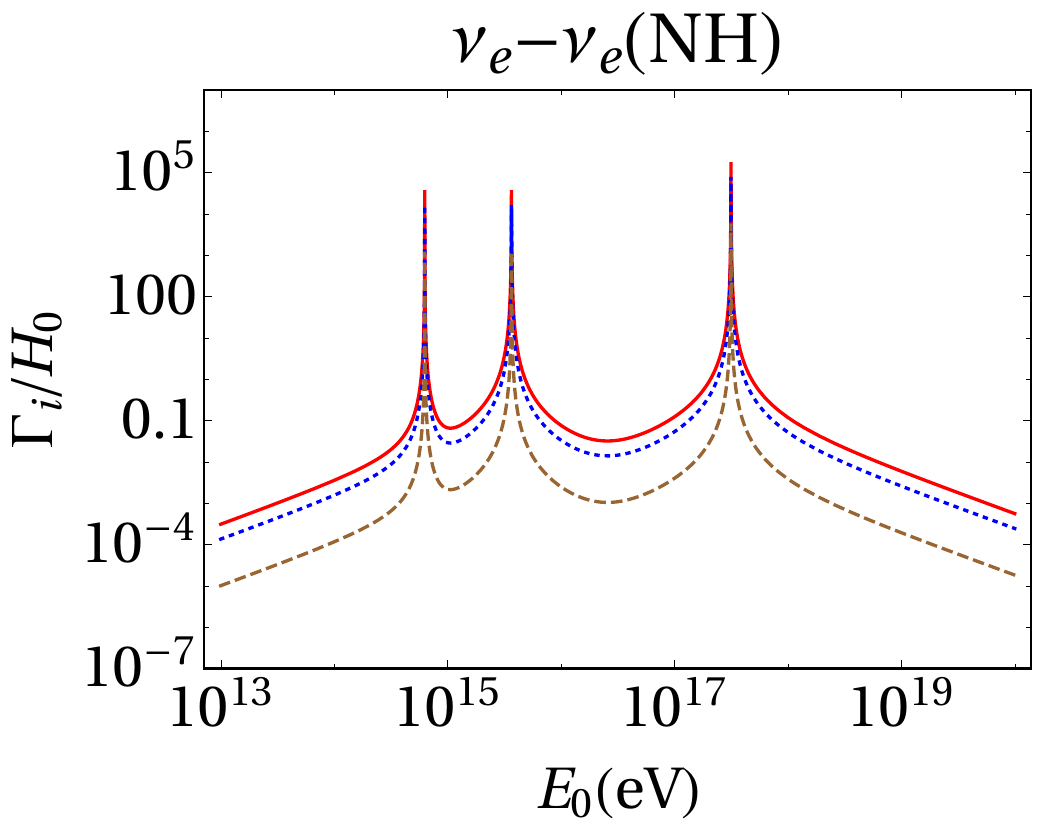}
\includegraphics[width=0.24\linewidth,height=0.2\linewidth]{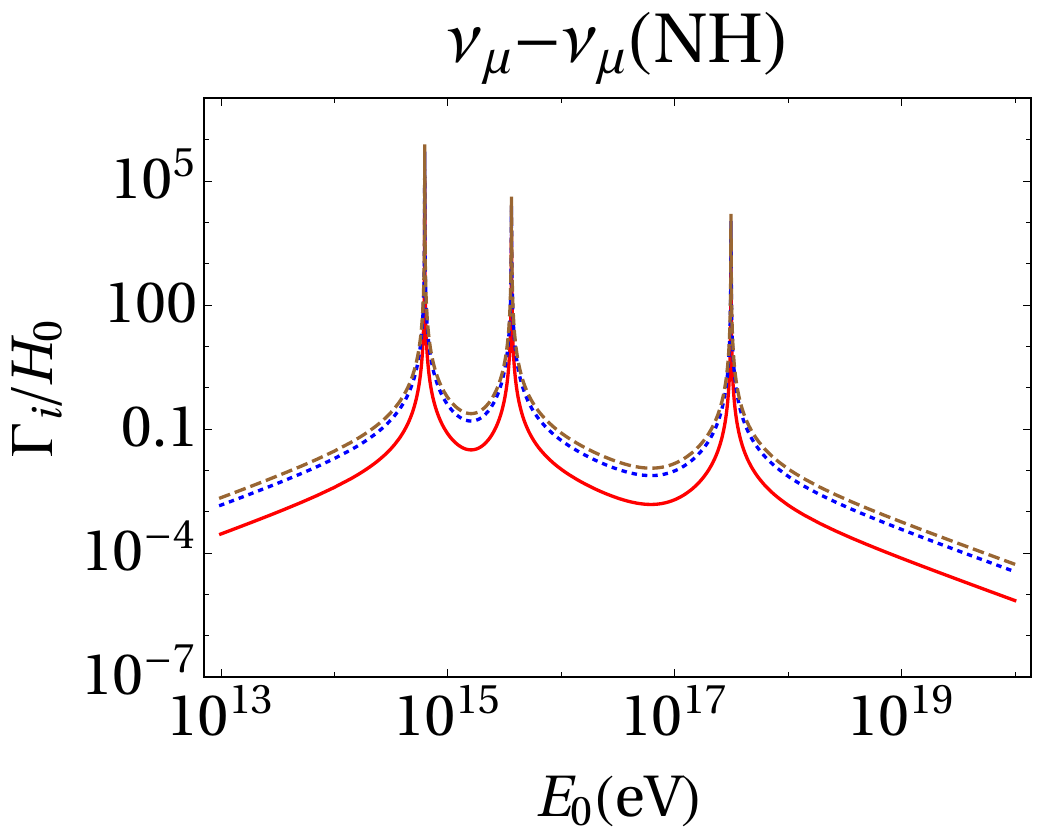}
 \includegraphics[width=0.24\linewidth,height=0.2\linewidth]{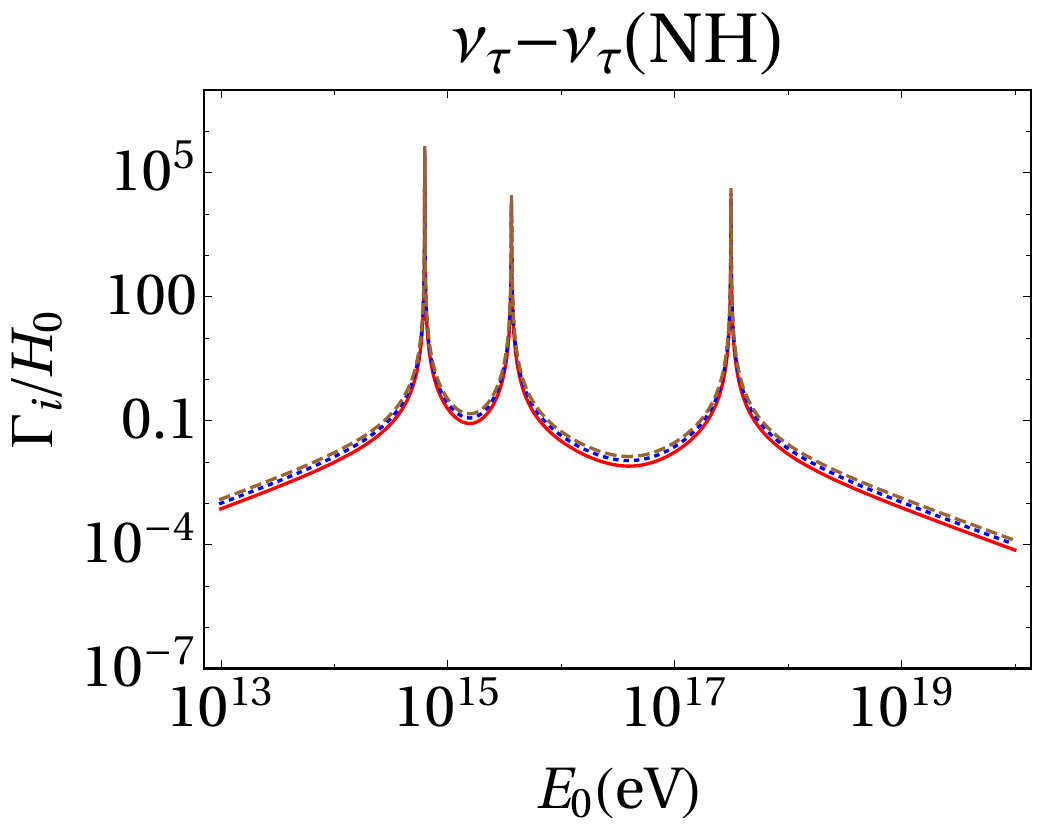}
\includegraphics[width=0.24\linewidth,height=0.2\linewidth]{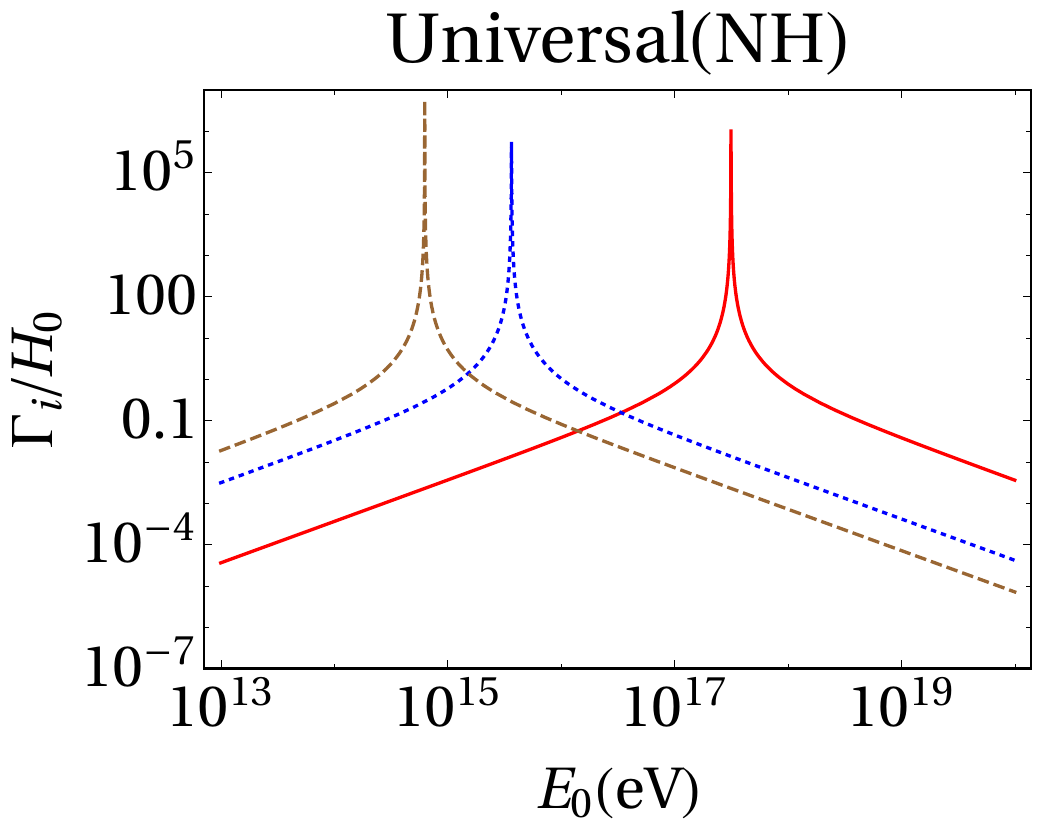}
 \includegraphics[width=0.24\linewidth,height=0.2\linewidth]{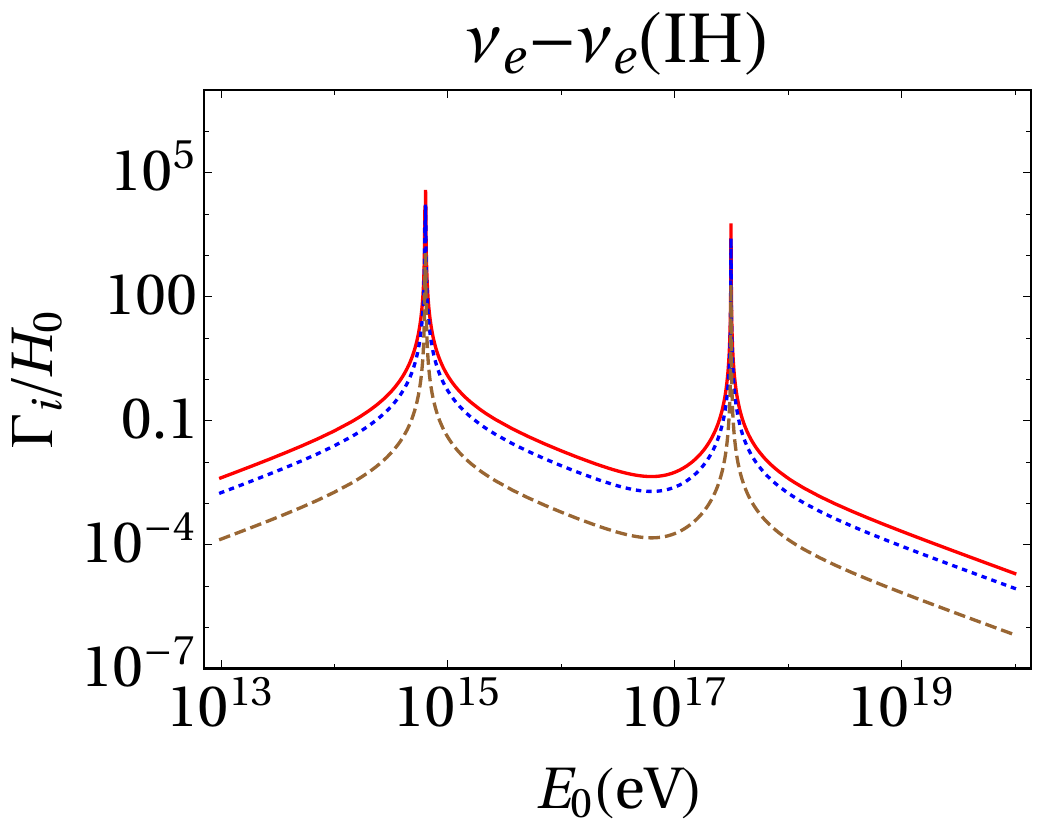}
\includegraphics[width=0.24\linewidth,height=0.2\linewidth]{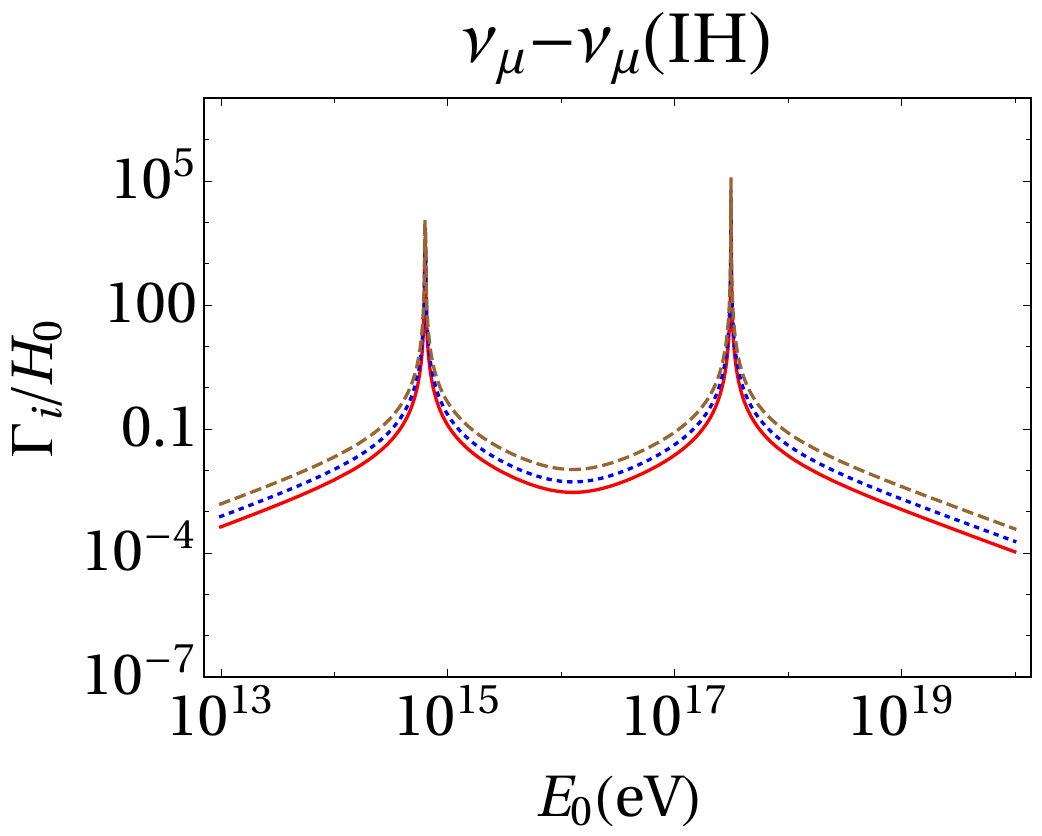}
 \includegraphics[width=0.24\linewidth,height=0.2\linewidth]{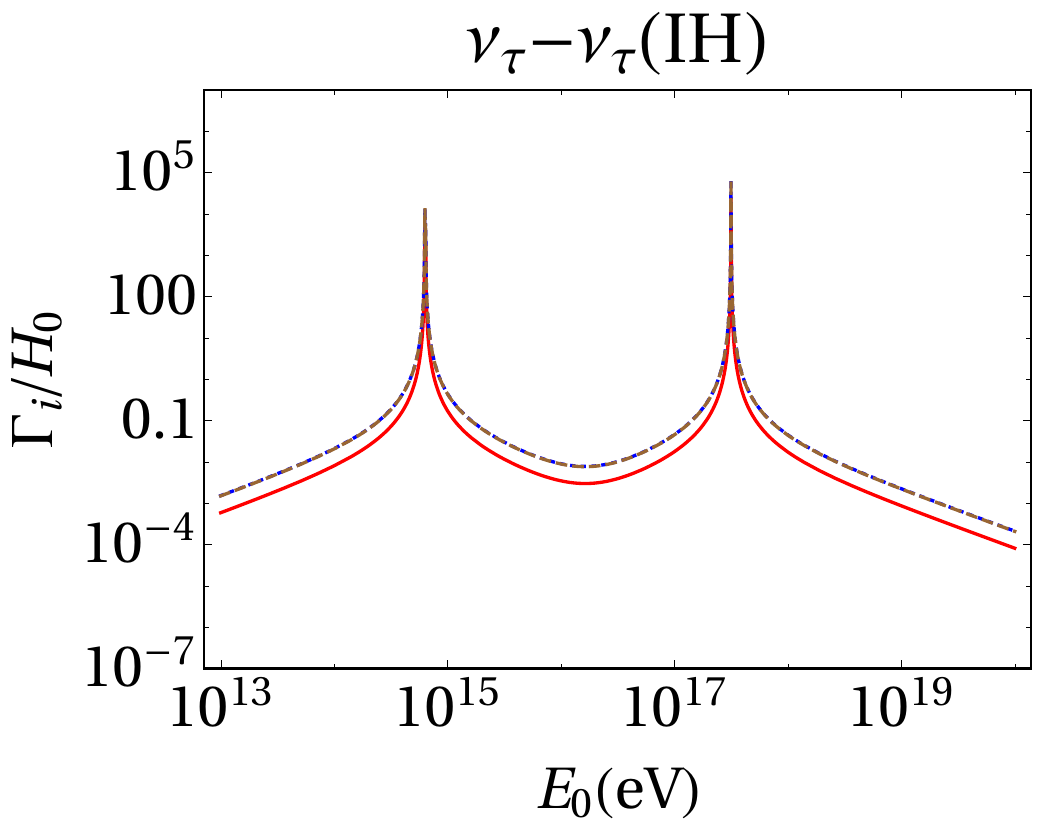}
\includegraphics[width=0.24\linewidth,height=0.2\linewidth]{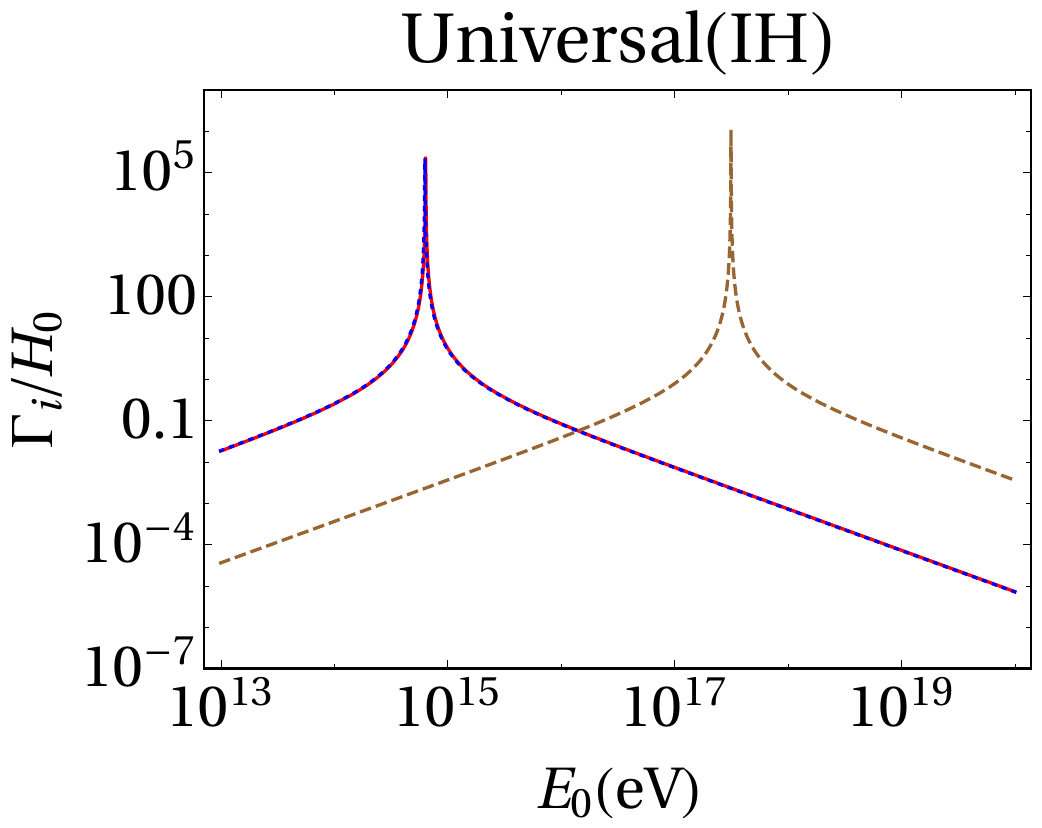}
\caption{Effects of different types of interaction on absorption rate ($\Gamma_i$) are shown. 
Red, blue (dotted) and brown (dashed) line demonstrate the interaction rates corresponding to mass eigenstates 
1,2 and 3 respectively. Here, the upper panel represents normal hierarchy and the lower panel represents inverted hierarchy. Values of the 
parameters for these plots kept fixed at $m_\phi=10^{6.9}$eV, $g_\phi=10^{-1.5}$ and $m_0=10^{-4}$eV.  }\label{gamma_compare}
\end{figure}

Propagation of the astrophysical neutrinos in the cosmic neutrino background can be described by the Boltzmann equation. The specific 
flux $\Phi_i$  of 
neutrino mass eigenstates $m_i$ is defined as 
\begin{eqnarray}
 \Phi_i = {\partial n_i\over\partial E}\, ,
\end{eqnarray}
where $n_i$ is the comoving number density of the astrophysical neutrinos per unit time and $E$ is the energy of the neutrino mass 
eigenstates. Therefore, the unit of the $\Phi_i$ is $\rm cm^{-2}s^{-1}sr^{-1}eV^{-1}$. The Boltzmann equation for $\Phi_i$ is defined  
as~\cite{Farzan:2014gza,Ng:2014pca} 
\begin{eqnarray}\label{eq:boltzmann_t}
 {\partial\Phi_i\over \partial t} = H\Phi_i + H E {\partial\Phi_i\over\partial E} + S_i(t,E) - \Gamma_i(t,E)\Phi_i + S_{{\rm 
tert},i}(t,E)\, .
\end{eqnarray}
Here, $H$ denotes the Hubble parameter, $S_i$ is the source term of the astrophysical neutrinos, $\Gamma_i$ is absorption rate and  $S_{{\rm 
tert},i}$ is the tertiary source term. Absorption rate $\Gamma_i = \sum_j \tilde{n}_j \sigma_{ij}$, where $\tilde{n}_j$ is  the comoving 
cosmological neutrino number density~\cite{Creque-Sarbinowski:2020qhz}.  Absorption rate is plotted against energy in 
\fig{gamma_compare}, we can see that $\Gamma_i$ becomes maximum at the resonance energies. 
For solving \eqn{eq:boltzmann_t}, we  recast it in redshift ($z$) variable as 
\begin{eqnarray}\label{eq:boltzmann_z}
 (1+z){\partial\Phi_i\over \partial z} + E {\partial\Phi_i\over\partial E} = -\Phi_i  - {S_i(z,E)\over H} + {\Gamma_i(z,E)\over H}\Phi_i - 
{S_{{\rm tert},i}(z,E)\over H}\, .
\end{eqnarray}
Solution of this equation is done using method of auxiliary equation~\cite{Farzan:2014gza,Ng:2014pca}, where the auxiliary equation is set 
to be
\begin{eqnarray}
 E = E_0(1+z)\, .
\end{eqnarray}
Here $E_0$ denotes the energy of neutrinos at $z=0$. The solution of \eqn{eq:boltzmann_z} can be written as~\cite{Ahlers:2009rf}
\begin{eqnarray}\label{eqn:phi}
 \Phi_i(E_0,z) = \int_0^\infty {dz'\over H(z')} \exp \left[\int_z^{z'} {dz''\over (1+z'')}{\Gamma_i(E,z'')\over H(z'')}\right]
 \lbrace S_i(E,z)+ S_{{\rm tert},i}(E,z)\rbrace\,.
\end{eqnarray}
As proposed in the literature, there can be many possible astrophysical sources for the high-energy neutrinos, which power the source term $S_i$. For example, gamma-ray blazars~\cite{Aartsen:2016lir, Hooper:2018wyk, Yuan:2019ucv, Luo:2020dxa, Smith:2020oac}, gamma-ray bursts~\cite{Waxman:1997ti, Abbasi:2009ig, Abbasi:2011qc, Abbasi:2012zw, Aartsen:2014aqy, Aartsen:2016qcr, Aartsen:2017wea}, Radio-bright AGN~\cite{Zhou:2021rhl},choked jet supernovae~\cite{Senno:2017vtd, Esmaili:2018wnv}, pulsar wind nebulae~\cite{Aartsen:2020eof}, etc. However, it is still unclear about the exact source of these high energy neutrinos. All these different sources can have different spectra for the neutrino flux. As an example, the source term related to the core-collapsed supernova (CCSN) reads as~\cite{Ando:2004hc} 
\begin{eqnarray}
 S_i(E,z) =  R_\text{CCSN}(z) {dN_i\over dE} E^{-\gamma_c}\, ,
\end{eqnarray}
where ${dN_i\over dE}$ is the comoving neutrino production rate per unit time per unit energy, and it is defined as 
\begin{eqnarray}
 {dN_i\over dE}=\frac{120 
E^2 E_{\rm tot}}{\left(7 \pi ^2\right) 6 (k_B T_{\rm sn})^4 \left(e^{E/k_B T_{\rm sn}}+1\right)}\, ,
\end{eqnarray}\label{eq:dndi}
where $ E_{\rm tot}=1.873\times10^{65}$eV and $ k_B T_{\rm sn}=8$ MeV~\cite{Creque-Sarbinowski:2020qhz}. It is evident from \eqn{eq:dndi} that the neutrino flux coming from CCSN will peak in the $\mathcal{O}(10)$ MeV range. Therefore, the CCSN can not produce a significant flux of ultra-high energy neutrinos relevant to the IceCube energy range. However, as argued in ref.\cite{Creque-Sarbinowski:2020qhz}, the redshift distribution of the ultra high energy neutrino sources is expected to closely follow the $R_\text{CCSN}(z)$. Therefore, we have considered the source term $S_i$ at the IceCube relevant energy range to be an effective power law as
\begin{eqnarray}
 S_i(E,z) \propto  R_\text{CCSN}(z) \Big(\frac{E}{E_0}\Big)^{-\gamma}\, .
\end{eqnarray}
Later in this section, we fix the normalization of the neutrino flux with the observed flux at 100 TeV.

$R_\text{CCSN}(z)$ represents the number density of the core-collapsed supernova as a function of $z$. We take this function from ref~\cite{Horiuchi:2008jz}   
\begin{eqnarray}
R_\text{CCSN}(z)=\dot{\rho} \left(\left(\frac{z+1}{B}\right)^{\beta  \eta }+\left(\frac{z+1}{C}\right)^{\gamma_2  \eta }+(z+1)^{\alpha  
\eta 
}\right)^{1/\eta } 
\end{eqnarray}
with the parameters being
$\dot{\rho}=0.0178$, $z_1=1$, $z_2=4$ $\alpha =3.4,\beta =-0.3,\gamma_2 =-3.5,\eta =-10$, $ B=(z_1+1)^{1-\frac{\alpha }{\beta }}$ and
$C=(z_2+1)^{1-\frac{\beta }{\gamma_2 }} (z_1+1)^{\frac{\beta -\alpha }{\gamma_2 }}$. 

Tertiary source term, ${S_{\rm tert,}}_i$ accounts for the up-scattering of the cosmological neutrinos from the collision with 
astrophysical neutrinos. This term is approximated in the literature in many different ways. In this paper we use the form provided by 
ref~\cite{Creque-Sarbinowski:2020qhz} which is 
\begin{eqnarray}
 {S_{\rm tert,}}_i(z,E) = \sum_{jkl}(1+\delta_{il})\tilde{n}_k(z) \sigma_{jkil}\Phi_j(z,E_{R_k})\Theta(E_{R_k} -E)\,.
\end{eqnarray}

\begin{figure}
 \centering
 \includegraphics[width=0.48\linewidth]{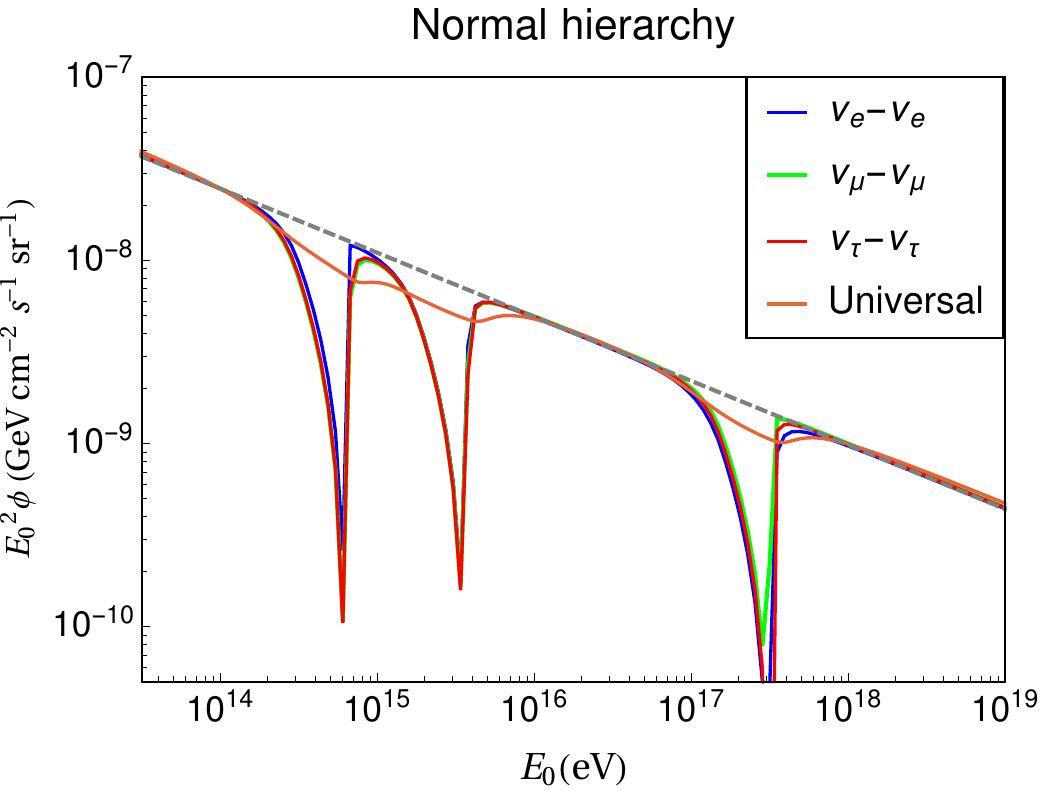}
 \includegraphics[width=0.48\linewidth]{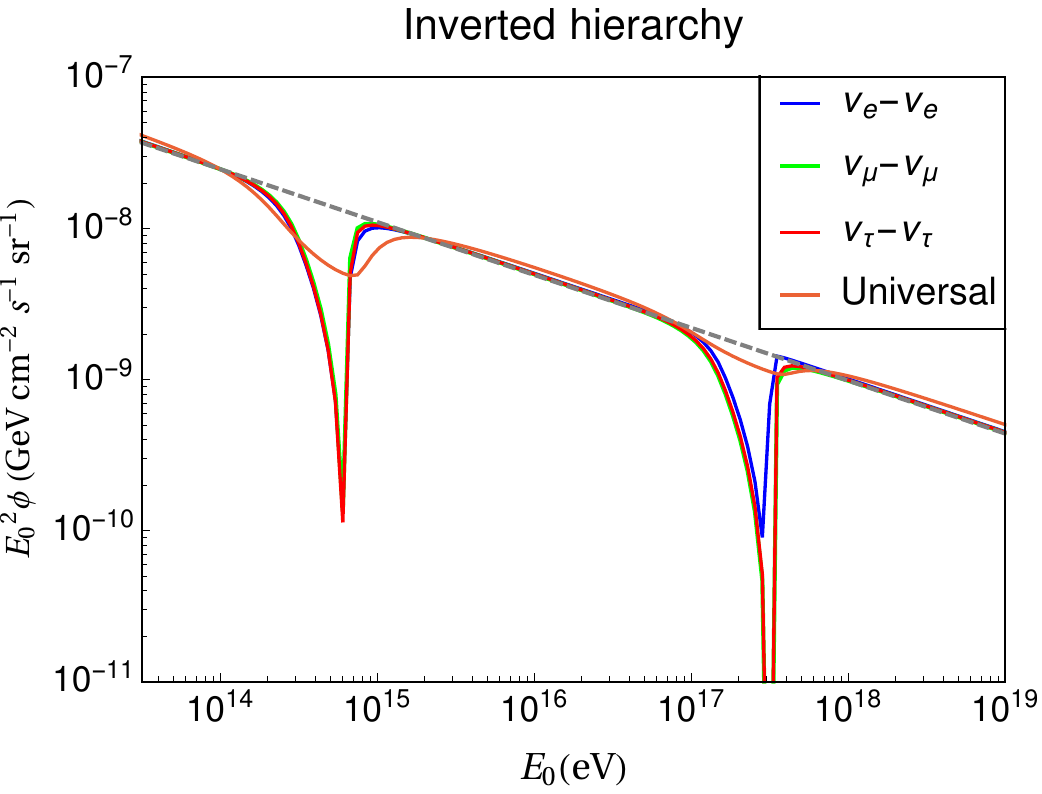}
 \caption{Effects of different types of self-interaction on total neutrino flux for both hierarchies are shown. Values of the parameters 
for these plots kept fixed at $m_\phi=10^{6.9}$eV, $g_\phi=10^{-1.5}$, $\gamma = 2.35$ and 
$m_0=10^{-4}$eV. Dashed line corresponds to $\Phi \propto E_0^{-\gamma}$. }\label{flux_compare}
\end{figure}

We have computed the specific flux of neutrino at $z=0$ by numerically solving \eqn{eqn:phi}. We have taken the maximum value of $z$ in 
this 
equation to be 10. It is because the $R_\text{CCSN}(z)$ function has non-negligible value up to redshift ten. Normalization of the $E_0^2 
\sum_i\Phi_i$ is 
fixed to $2.46 \times 10^{-8}{\rm GeV\,cm^{-2}s^{-1}sr^{-1}}$  for neutrino energy equal to 100 TeV~\cite{Kopper:2017zzm}. The results for 
different types of interactions and hierarchies are 
shown in \fig{flux_compare}. To understand these plots we have made another set of plots in \fig{gamma_compare} where the values of 
absorption rate, $\Gamma_i$ has been plotted for different interactions. We see that there is a major difference between the universal 
interaction and the flavour specific interactions. In case of universal interaction the $g_{ij}$ and $g_{kl}$ in \eqn{eqn:sig4} becomes 
kronekar delta function and thus $\Gamma_i$ gets the contribution from $s_i$ only. However, for flavour specific interactions $g_{ij}$ 
mixes 
all the mass eigenstates and $s_j$ corresponding to all mass eigenstates contributes in $\Gamma_i$. Therefore, in \fig{gamma_compare} we 
see that for flavour specific interactions the $\Gamma_i$ shows resonance peaks in all three possible energy values corresponding to the 
three mass eigenstates in normal hierarchy. In the case of inverted hierarchy the mass gap between first two mass eigenstates are 
small, and the peaks corresponding to those mass eigenstates are indistinguishable, therefore, only two peaks are separately visible. These 
peaks in 
$\Gamma_i$ leads to the absorption dips in the flux of astrophysical neutrinos in \fig{flux_compare}. Since, in the case of flavour 
specific 
interactions all mass eigenstates undergo absorption in all dips (three dips for NH and two dips for IH), we see that the total flux also 
shows dips in all the resonance energies.  However, in the case universal interaction in which energy one mass 
eigenstate undergoes resonant absorption other mass eigenstates do not. Therefore, in \fig{flux_compare} the universal curve shows a more 
flat 
line. The three different peaks in the $\Gamma_i$ of universal interaction can come close together if neutrino masses become degenerate. In 
that 
case a prominent dip will be visible in the total neutrino flux for universal interaction also. However, that means it requires a higher 
value 
lowest neutrino mass and in the next section we will check if that is consistent with CMB bound.

We found that the contribution of the tertiary source term compared to the core-collapse supernova source term is negligibly small. 
Moreover, it considerably increases the computation time. Therefore, for constraining the parameter space using \ic data we neglect the 
tertiary term in the next section.

\section{Parameter estimation from flux at IceCube}\label{IceCube}
\begin{figure}[h]\centering
 \includegraphics[width=0.5\linewidth,height=0.32\linewidth]{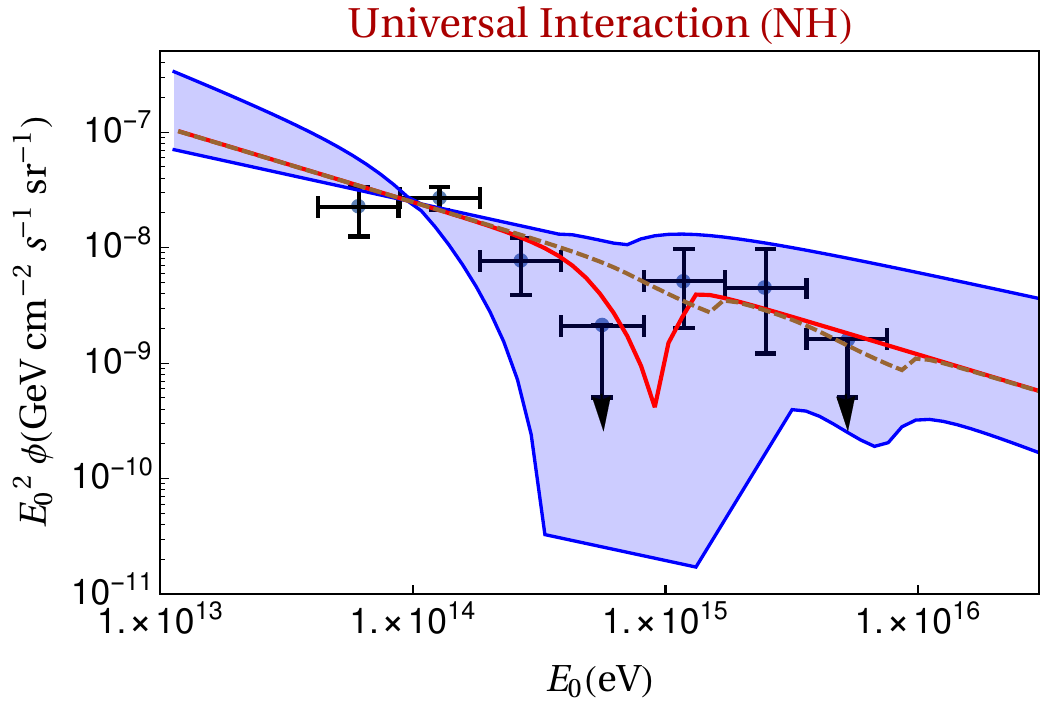}
 \includegraphics[width=0.45\linewidth,height=0.32\linewidth]{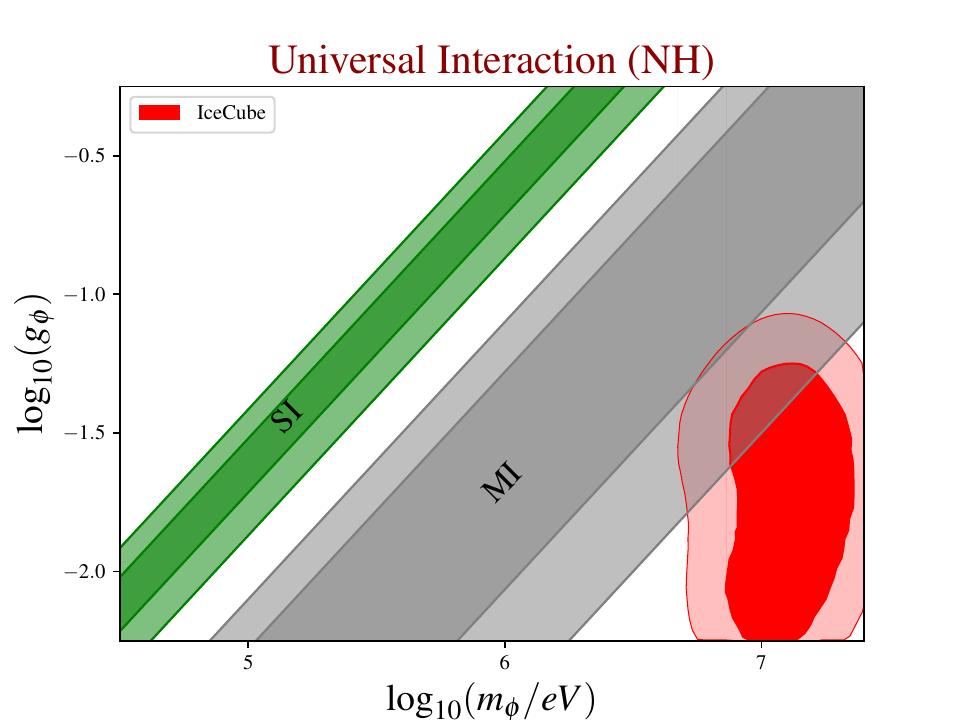}\\
 (a) \hspace*{0.45\linewidth} (b)
 \caption{{\bf Universal interaction in normal hierarchy:} (a) The shaded region corresponds to the specific flux of the neutrinos allowed by the 1-$\sigma$ 
ranges of all the parameters ($m_\phi,g_\phi$, $\gamma$ and $m_0$) as given in table~\ref{tab:norm}. The solid red line corresponds to the 
bestfit values of all these parameters. However, the orange dashed line corresponds to parameters $m_\phi,g_\phi$ and $\gamma$ fixed at their 
bestfit values and $m_0$ value fixed at zero. (b) The allowed 1-$\sigma$ and 2-$\sigma$ regions in the $m_\phi$-$g_\phi$ 
plane from \ic data are shown along with the cosmological bound (``Planck+BAO+HST'') for universal interaction in the normal hierarchy. The allowed region for \ic has no overlap with the strong self-interaction (SI) band.  
}\label{fig:uni_norm_ic}
\end{figure}

In IceCube six-year HESE data, 82 events passed the selection criterion of which 
two are coincident with atmospheric muons and left out. The best fit for single power law flux is~\cite{Kopper:2017zzm}
\begin{equation}
E_0^2 \phi = (2.46 \pm 0.8) \times 10^{-8} 
\left( \frac{E_0}{100~{\rm TeV}}
\right)^{-0.92} 
 ~ {\rm GeV\,cm^{-2}s^{-1}sr^{-1}}\, ,
\end{equation}
which has a softer spectral index than the 3-year ($\gamma = 2.3 $) as well as the 4-year ($\gamma = 2.58 $) data. These events 
are binned in 6 values of energy and for other values of energy where no events have been observed, an upper bounds on the 
neutrino flux have been provided. For fitting our numerical results of flux with the \ic data we have solved \eqn{eqn:phi}
for 63 different values of $E_0$ which are equally spaced in log scale within $E_0$-range of first 7 points in \ic data (see
\fig{fig:uni_norm_ic}). Therefore, we have sampled each energy bin with 9 points in the log scale. The average of these 9 points 
are assigned as the theoretical prediction of neutrino flux. These theoretical predictions are then compared to the observational 
values of binned flux and the allowed parameter space ($m_\phi$, $g_\phi$, $m_0$ and $\gamma$) has been estimated using the MCMC technique with 
Metropolis-Hastings algorithm. Please refer to the 
appendix~\ref{app:numerical} for details.

Before moving forward to describe our findings for different types of interactions and hierarchies, let us briefly summarize the effects of 
different parameters on the features of the specific flux of neutrinos in \ic. These effects can be listed as
\begin{itemize}
 \item The higher value of $m_\phi$ shifts the dips towards the higher values of neutrino energy.
 \item The higher values of lowest neutrino mass $m_0$ make the difference between the neutrino masses smaller and thus make 
absorption dips come closer.
\item In case of universal interaction higher $m_0$ makes dips sharper and lower $m_0$ makes the flux more flat.
\item In general, higher values of $m_0$ make the neutrinos heavier and therefore the dips move towards lower values of neutrino energy.
\item Higher values of $g_\phi$ make the dips sharper.
\item $\gamma$ determines the slope of the flux line. The more $\gamma$ becomes close to the value 2 the more the flux line becomes 
flat. 
\end{itemize}
In the next subsections we will discuss our findings of MCMC analysis of parameter space for both the hierarchies and different types of 
interactions.
\subsection{Normal hierarchy}
\begin{figure}\centering
  \includegraphics[width=0.45\linewidth]{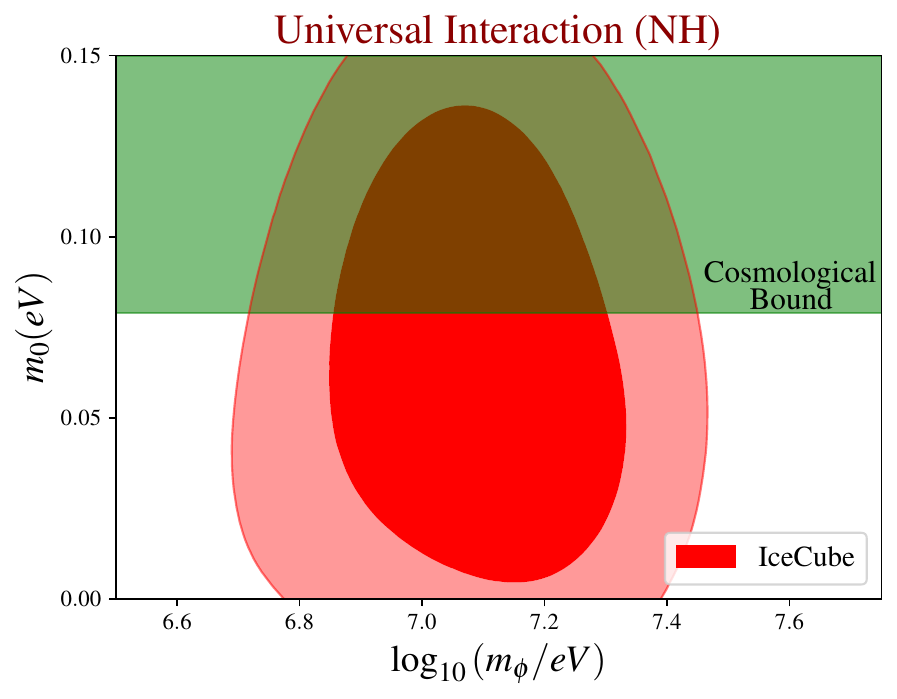}
   \includegraphics[width=0.45\linewidth]{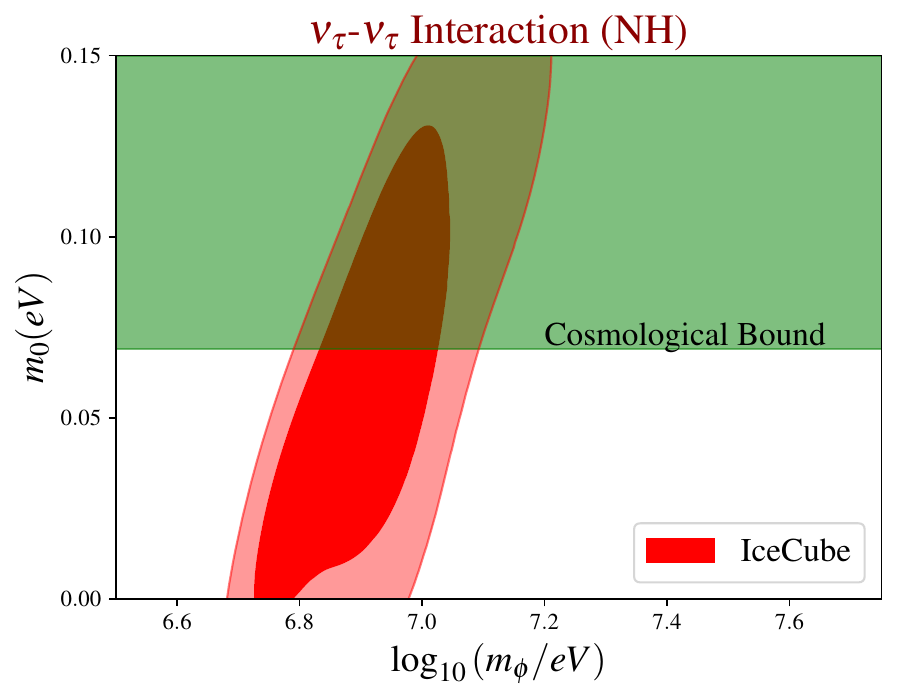}\\
   (a)Universal interaction \hspace*{2.3in}(b)$\nu_\tau$-$\nu_\tau$ interaction
\caption{The green region shows the values of lowest neutrino mass ($m_0$) excluded by cosmological data (``Planck+BAO+HST'') for the normal 
hierarchy. A major portion of the preferred $m_0$ values by \ic data is disfavored by the cosmological bound.  
}\label{cmb_mass_bound_normal}
\end{figure}

\paragraph{Universal interactions:}
Number of visible dips in the neutrino specific flux heavily depends on the value of the lowest neutrino mass $m_0$. 
Moreover, as shown in the 
\fig{flux_compare} the universal interaction produces more flat line of specific flux compared to the flavour specific 
interactions for smaller values of $m_0$. However, when the $m_0$ is large, the absorption dips in flux lines corresponding 
to the different mass eigen states come closer and the dips in the total flux become sharper.  Therefore, to fit the dips in the \ic data 
the universal interaction prefers a higher value of $m_0$.  The bestfit values along with 1-$\sigma$ error for all the parameters 
($m_\phi,g_\phi,\gamma$ amd $m_0$) are given in table~\ref{tab:norm}. Moreover, the cosmological 1-$\sigma$ upper bound of $m_0$ for the 
case of moderate universal interactions in the normal hierarchy is 0.079 eV (see table~\ref{tab:mi_values}). 
 The disallowed values of $m_0$ are shown as the green shaded region in \fig{cmb_mass_bound_normal}-(a). Therefore, It is clear from 
\fig{cmb_mass_bound_normal}-(a) that a substantial region of the preferred mass range by \ic is disallowed by the cosmological bound.

In \fig{fig:uni_norm_ic}, the specific flux in \ic and the corresponding allowed parameter space for $m_\phi$-$g_\phi$ have been shown.  
The shaded region in the \fig{fig:uni_norm_ic}-(a) corresponds to the specific flux of the neutrinos by the maximum allowed values of all
the parameters ($m_\phi,g_\phi,\gamma$ amd $m_0$) within the 1-$\sigma$ range. The shaded 
region shows that the dip in the specific flux can reach only up to a certain minima and it cannot go deeper since that will require larger 
neutrino masses than the specified upper bound on $m_0$ in table~\ref{tab:norm}.    
We have also plotted a red solid line and a dashed orange line of the specific flux. The red solid line corresponds to bestfit values of all 
the parameters. Whereas, the orange line corresponds to the $m_0=0$ eV and all other parameters $m_0$, $g_\phi$ and $\gamma$ fixed to their 
best-fit values. As discussed in the previous section, ``Planck+BAO+HST'' data also puts a bound on $G_{\rm eff}$, which 
translate into a bound on $m_\phi$-$g_\phi$ parameter space. The 1-$\sigma$ and 2-$\sigma$ bounds for both SI and MI region have been shown 
by the 
green and gray shaded area respectively in \fig{fig:uni_norm_ic}-(b). It is quite evident from \fig{fig:uni_norm_ic}-(b) that the \ic 
allowed parameter space for $m_\phi$-$g_\phi$ is inconsistent with SI bound. Whereas, some part of \ic allowed parameter space for 
$m_\phi$-$g_\phi$ overlaps with MI bound at 2-$\sigma$ level only. 

\paragraph{$\nu_\tau$-$\nu_\tau$ interactions:}
  \begin{figure}\centering
 \includegraphics[width=0.45\linewidth,height=0.32\linewidth]{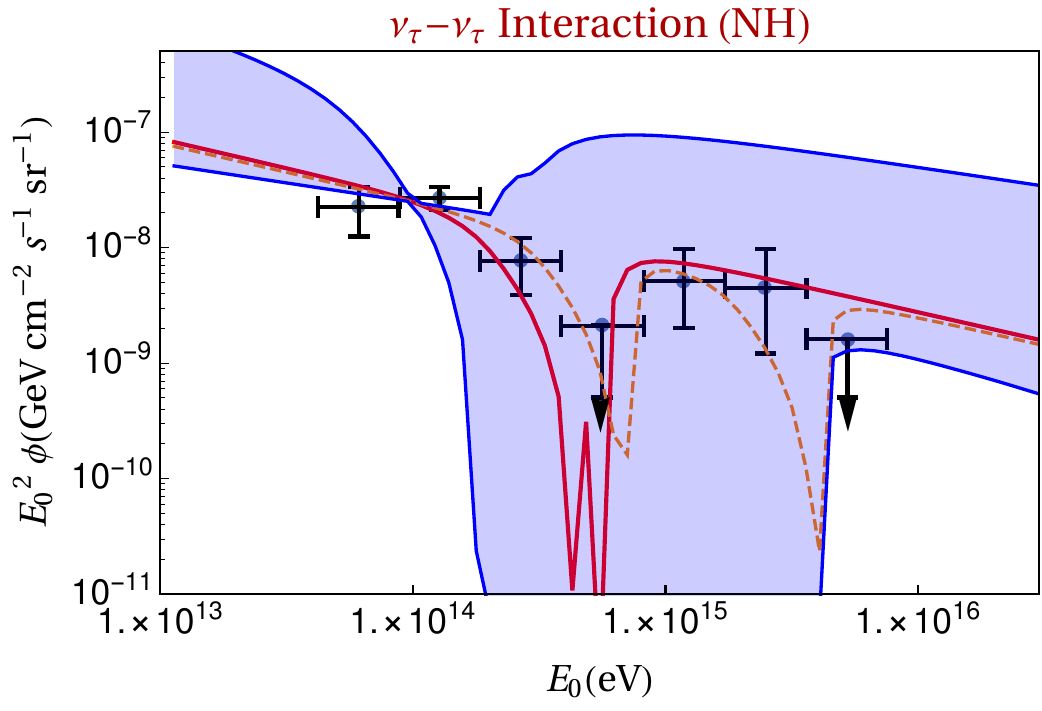}
 \includegraphics[width=0.45\linewidth,height=0.32\linewidth]{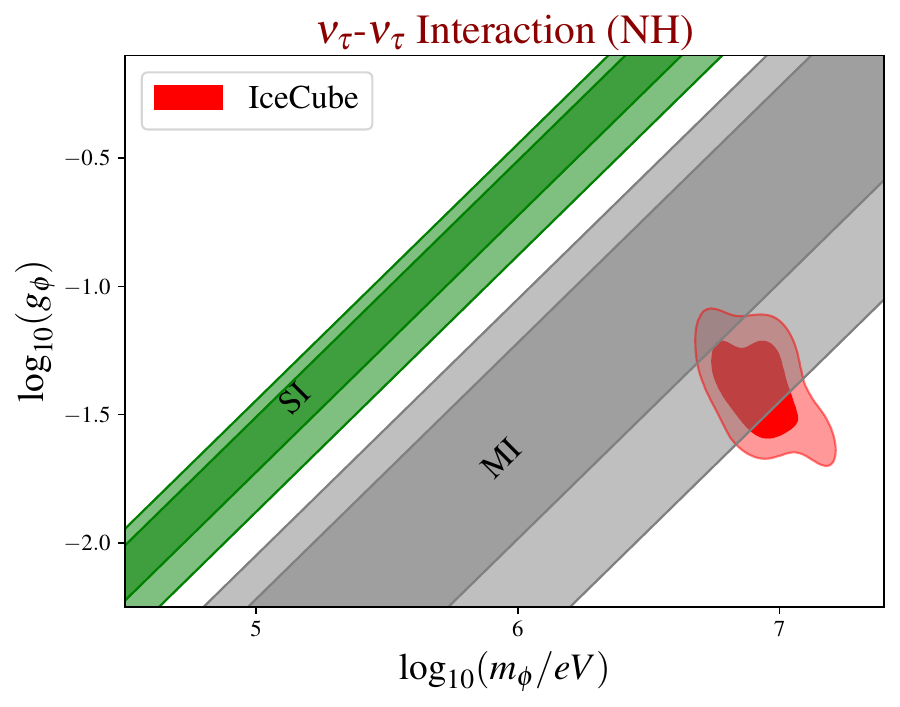}\\
 (a) \hspace*{0.45\linewidth} (b)
\caption{{\bf $\nu_\tau$-$\nu_\tau$ interaction in normal hierarchy:} (a) The shaded region corresponds to the specific flux of the neutrinos allowed by the 1-$\sigma$ 
ranges of all the parameters in table~\ref{tab:norm}. The solid red line correspond to the best fit values of all the 
parameters and 
the orange dashed line  corresponds to the best fit value of $m_\phi,g_\phi$ and $\gamma$ and the value of $m_0$ 
is fixed at zero. (b) The allowed 1-$\sigma$ and 2-$\sigma$ regions in the $m_\phi$-$g_\phi$ 
plane from \ic data are shown along with the bounds from the cosmological bound (``Planck+BAO+HST'') for $\nu_\tau$-$\nu_\tau$ interaction 
in the normal hierarchy. The 
allowed region for \ic has no overlap with the strong self-interaction (SI) band.}\label{fig:tau_norm_ic}
 \end{figure}
The main difference in the features of flavour specific interaction and universal interaction is that while flavour specific interaction 
can produce two prominent dips in two different energy bins of \ic data, the universal interaction allows only one 
prominent 
dip. However, the two dips in a flavour specific interaction can be merged to one by increasing the neutrino mass.
The bestfit values along with 1-$\sigma$ error for all the parameters ($m_\phi,g_\phi,\gamma$ amd $m_0$) is given in table~\ref{tab:norm}. The 
specific flux of the neutrinos corresponds to the 1-$\sigma$ allowed values of all these parameters is shown by the shaded region in the 
\fig{fig:tau_norm_ic}-(a).
 We have also plotted two different lines, the solid red and orange dashed line, for specific flux in \ic which 
show quite different features.
For the orange dashed line, $m_0$ has been taken to be zero  and all other parameters has been kept at 
their best fit values. This line can explain two dips in \ic data at two different energies. If the neutrino mass is increased, the orange 
dashed line in the figure moves towards the solid red line which corresponds to the bestfit values of all the parameters presented in 
table~\ref{tab:norm}.     
Therefore we see in the \fig{fig:tau_norm_ic}-(a) that the solid red line and the orange dashed line represent the one and two dip 
solutions in the \ic energy range. Similar to the universal case, it is quite evident from \fig{fig:tau_norm_ic}-(b) that the \ic allowed 
parameter space for $m_\phi$-$g_\phi$ is inconsistent with SI region. Whereas, \ic allowed parameter space for $m_\phi$-$g_\phi$ is 
consistent with MI bound at 2-$\sigma$ level. We have also presented the \ic allowed parameter space for $m_\phi$-$m_0$ in 
\fig{cmb_mass_bound_normal}-(b). The cosmological 1-$\sigma$ upper bound of $m_0$ is 0.069 eV for moderate $\nu_\tau$-$\nu_\tau$ 
interaction in case of normal hierarchy. The disallowed region is shown in the green color in \fig{cmb_mass_bound_normal}-(b), which rules 
out a substantial part of \ic allowed $m_0$ values.

\begin{table}\centering
\begin{tabular}{| l  c|}\hline
Parameter &  68\% limits\\
\hline
{\bf \underline{Universal interaction}}&\\
{\boldmath$m_0 ({\rm eV})            $} & $0.067^{+0.038}_{-0.046}   $\\

{\boldmath$\log_{10}({m_\phi/{\rm eV}})      $} & $7.09\pm 0.15              $\\

{\boldmath$\log_{10}{g_\phi}        $} & $-1.75\pm 0.31             $\\

{\boldmath$\gamma         $} & $2.66^{+0.21}_{-0.18}      $\\
\hline
{\bf \underline{$\nu_\tau$-$\nu_\tau$ interaction}}&\\
{\boldmath$m_0  ({\rm eV})         $} & $0.062^{+0.042}_{-0.046}   $\\

{\boldmath$\log_{10}{m_\phi/{\rm eV}}       $} & $6.92\pm 0.11              $\\

{\boldmath$\log_{10}{g_\phi}       $} & $-1.40\pm 0.13             $\\

{\boldmath$\gamma         $} & $2.50\pm 0.17              $\\
\hline
\end{tabular}\caption{The preferred bestfit values of the \ic parameters and their 1-$\sigma$ ranges for both universal and $\nu_\tau$-$\nu_\tau$ 
interaction case in the {\bf normal hierarchy.} are listed in this table}\label{tab:norm}
\end{table}
 

%
%
%

%
%

\subsection{Inverted Hierarchy}
\begin{figure}\centering
 \includegraphics[width=0.45\linewidth,height=0.32\linewidth]{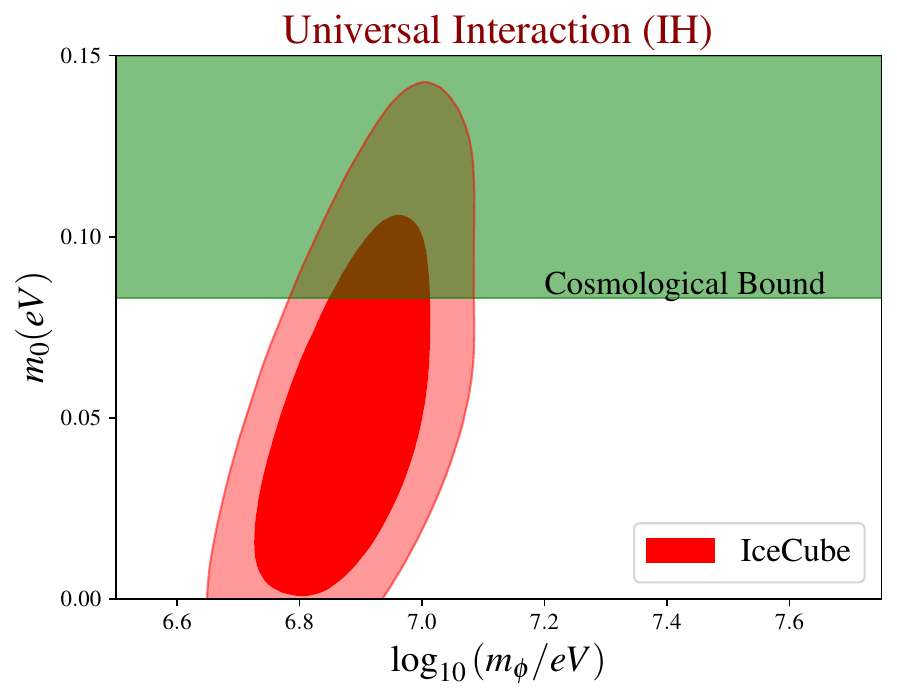} 
 \includegraphics[width=0.5\linewidth,height=0.32\linewidth]{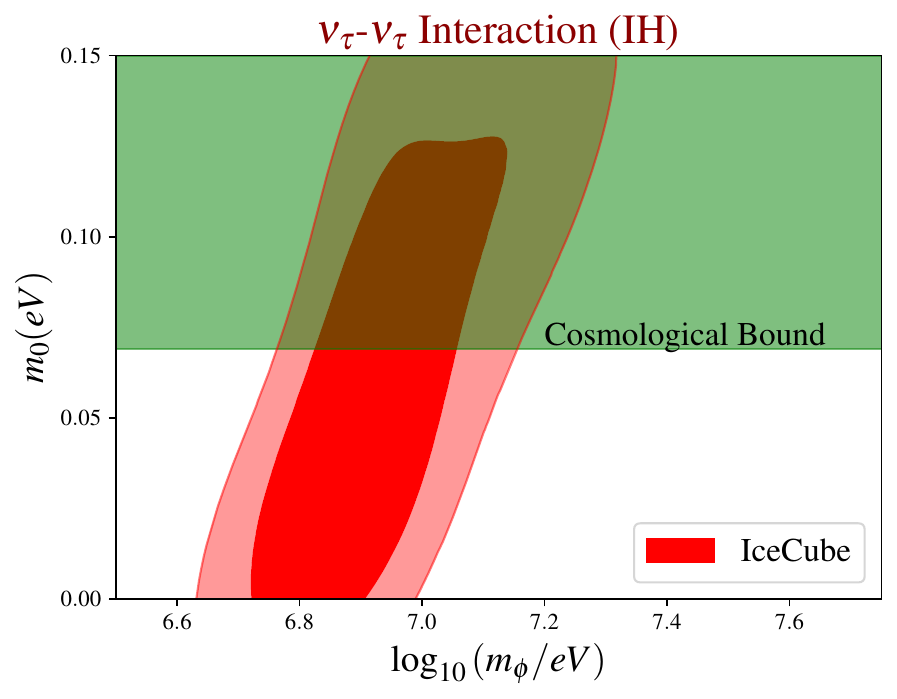}\\
 (a) Universal interaction \hspace*{3.2in}(b) $\nu_\tau$-$\nu_\tau$ interaction
 \caption{The green region shows the values of lowest neutrino mass ($m_0$) excluded by cosmological data (``Planck+BAO+HST'') for the inverted hierarchy.  For both the interaction cases, a significant portion of the preferred $m_0$ values by \ic data is excluded by the cosmological bound.}\label{cmb_mass_bound_inverted}
\end{figure}

\paragraph{Universal interactions:}
 Similar to the case of normal hierarchy, in the case of inverted hierarchy, the universal interaction produces a flatter specific 
flux line compared to the flavour specific 
interactions for smaller values of $m_0$  as well. Since large values of $m_0$ make masses of different neutrino mass eigenstates 
almost degenerate, we get a single dip solution for such case. Therefore, again to fit the dips in the \ic data the universal interaction 
prefers a higher value of $m_0$. The bestfit values along with 1-$\sigma$ error for all the parameters ($m_\phi,g_\phi,\gamma$ amd $m_0$) are 
given in table~\ref{tab:inv}. ``Planck+BAO+HST" data also put a bound on the lowest neutrino mass $m_0 \le 
0.083$ eV for the moderate universal interaction in the inverted hierarchy. The values of $m_0$ excluded by the cosmological data are shown in the green shaded region in \fig{cmb_mass_bound_inverted}, which clearly implies that  a substantial fraction of the \ic preferred mass range is disallowed by the cosmological bound. 

\begin{figure}\centering
 \includegraphics[width=0.45\linewidth,height=0.32\linewidth]{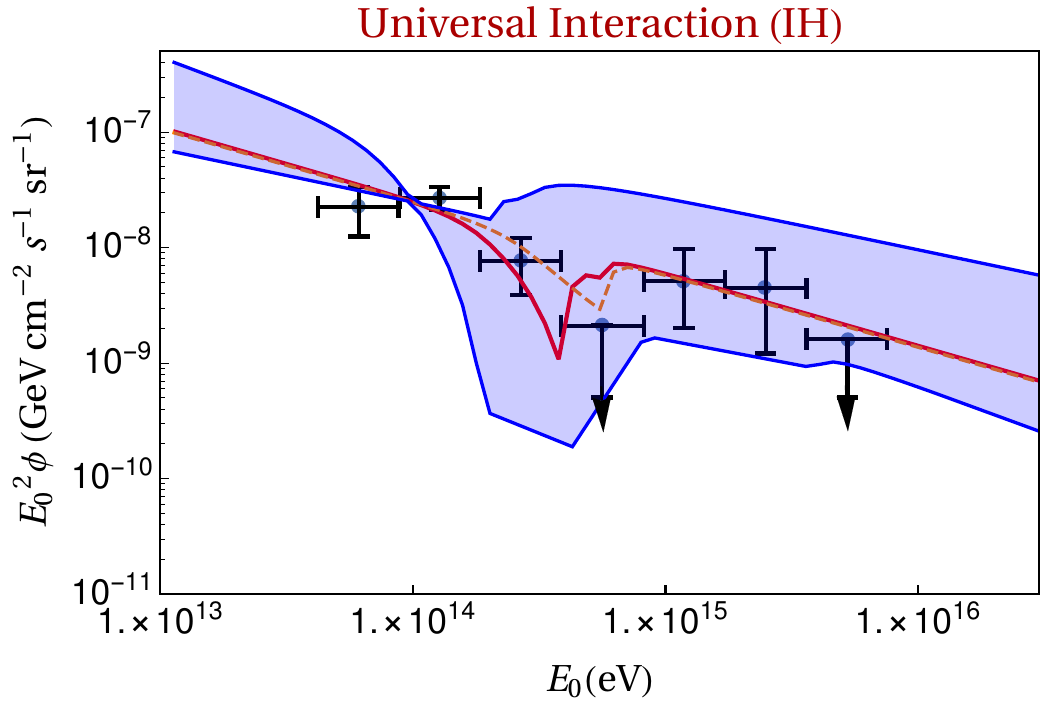}
 \includegraphics[width=0.45\linewidth,height=0.32\linewidth]{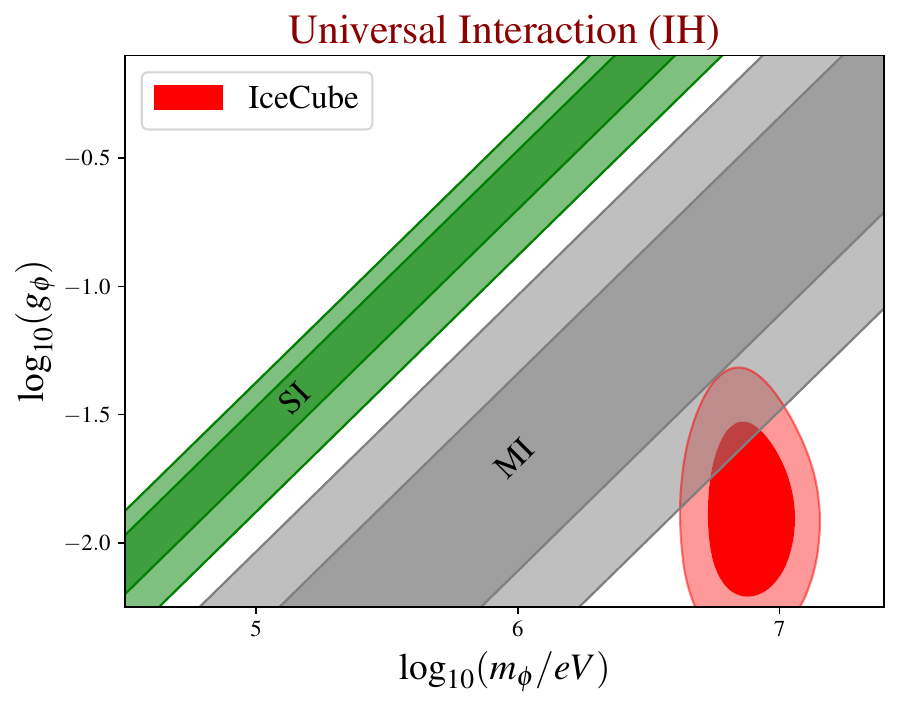}\\
 (a) \hspace*{0.45\linewidth} (b)
\caption{{\bf Universal interaction in inverted hierarchy:} (a) The shaded region corresponds to the specific flux of the neutrinos allowed by the 1-$\sigma$ 
ranges of the parameters in table~\ref{tab:inv}. The solid red line correspond to the best fit values of the parameters and the orange 
dashed line  corresponds to the best fit value of $m_\phi,g_\phi$ and $\gamma$ (in table~\ref{tab:inv}) and the value of $m_0$ is 
fixed at zero. (b) The allowed 1-$\sigma$ and 2-$\sigma$ regions in the $m_\phi$-$g_\phi$ 
plane from \ic data are shown along with the cosmological bound (``Planck+BAO+HST'') for universal interaction in the inverted hierarchy. The allowed 
region for \ic has no overlap with the strong self-interaction (SI) band.}\label{fig:uni_inv_ic}
 \end{figure}

In \fig{fig:uni_inv_ic}, the specific flux in \ic and the corresponding allowed parameter space for $m_\phi$-$g_\phi$ have been shown. The 
shaded region in the \fig{fig:uni_inv_ic}-(a) corresponds to the specific flux of the neutrinos for the maximum allowed values of all the 
parameters ($m_\phi,g_\phi,\gamma$ and $m_0$) within the 1-$\sigma$ range. Similar to 
the normal hierarchy case,  the shaded 
region shows that the dip in the specific flux can reach only up to a certain minima and it cannot go deeper cause that will require 
larger neutrino masses.    
We have also plotted a red solid line and a dashed orange line of specific flux. The red solid line corresponds to the best fit values of 
all the parameters listed in table \ref{tab:inv}. Whereas, the orange line corresponds to the $m_0=0$ eV and all other parameters fixed at 
their best fit values. Similar to the normal hierarchy, It is quite evident from \fig{fig:uni_inv_ic}-(b) that the \ic allowed parameter 
space for $m_\phi$-$g_\phi$ is inconsistent with SI bound. However, some part of \ic allowed parameter space for $m_\phi$-$g_\phi$ overlaps 
with MI bound at 2-$\sigma$ level only.

\begin{table}[b]\centering
\begin{tabular}{| l  c|}\hline
Parameter &  68\% limits\\
\hline
{\bf \underline{Universal interaction}}&\\
{\boldmath$m_0  ({\rm eV})       $} & $0.052^{+0.034}_{-0.040}   $\\

{\boldmath$\log_{10}({m_\phi/{\rm eV}})       $} & $6.883\pm 0.091            $\\

{\boldmath$\log_{10}{g_\phi}      $} & $-1.87\pm 0.20             $\\

{\boldmath$\gamma         $} & $2.63\pm 0.17              $\\
\hline
{\bf \underline{$\nu_\tau$-$\nu_\tau$ interaction}}&\\
{\boldmath$m_0  ({\rm eV})        $} & $0.062^{+0.038}_{-0.058}   $\\

{\boldmath$\log_{10}({m_\phi/{\rm eV}})      $} & $6.936^{+0.089}_{-0.14}    $\\

{\boldmath$\log_{10}{g_{\phi}}      $} & $-1.25\pm 0.17             $\\

{\boldmath$\gamma         $} & $2.46^{+0.18}_{-0.16}      $\\
\hline
\end{tabular}\caption{The preferred bestfit values of the \ic parameters and their 1-$\sigma$ ranges for both universal and $\nu_\tau$-$\nu_\tau$ 
interaction case in the {\bf inverted hierarchy.} are listed in this table.}\label{tab:inv}
\end{table}

\begin{figure}\centering
 \includegraphics[width=0.45\linewidth,height=0.32\linewidth]{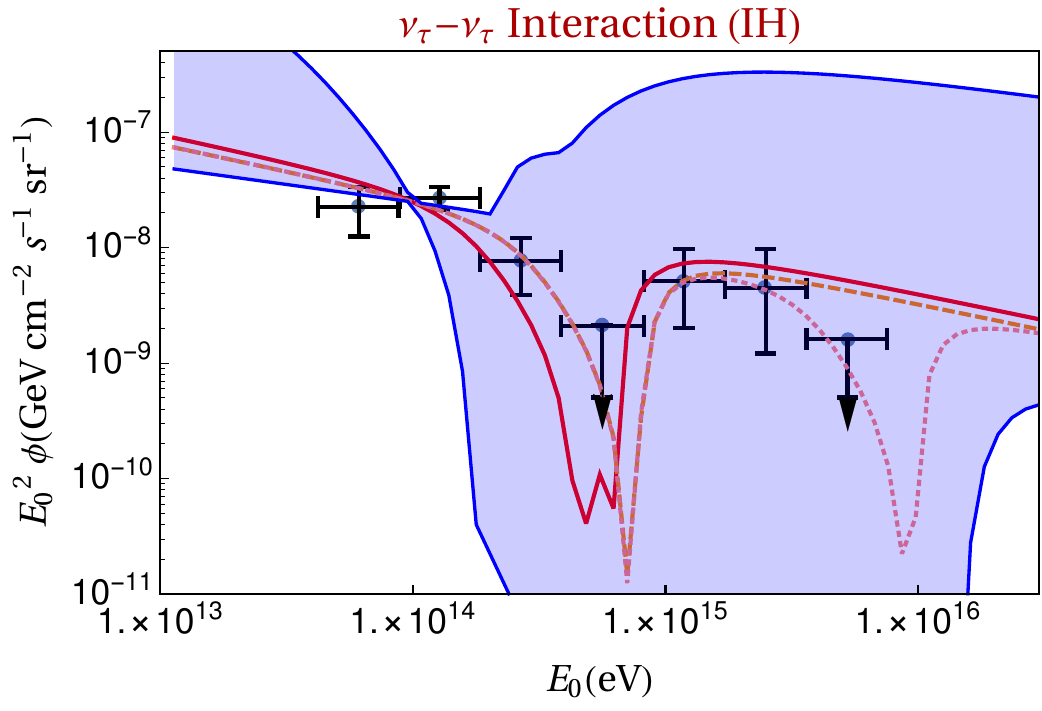}
 \includegraphics[width=0.45\linewidth,height=0.32\linewidth]{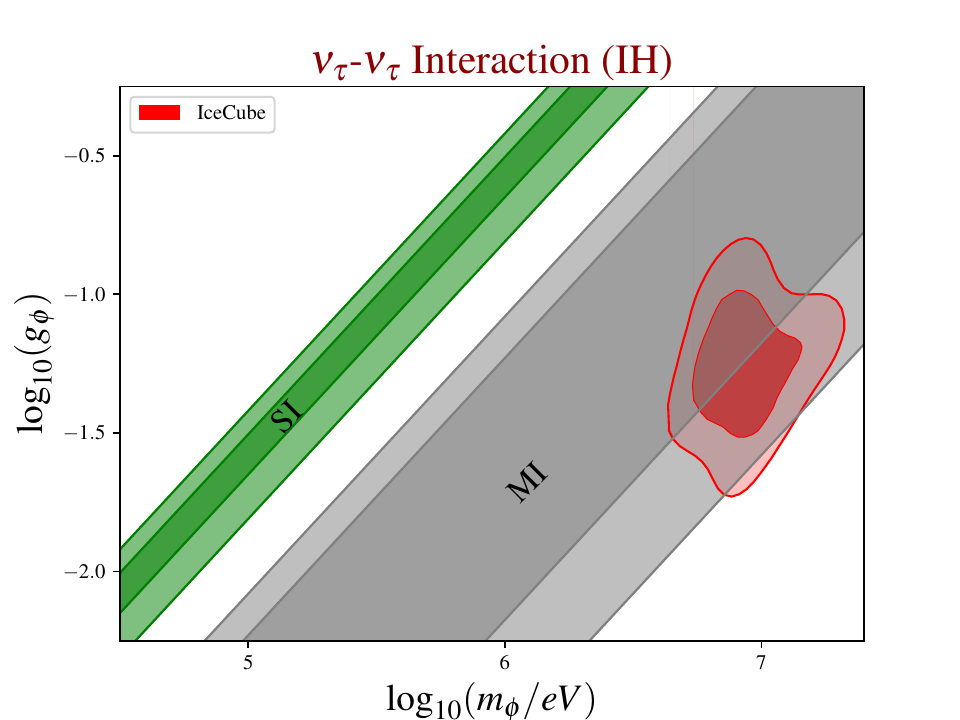}\\
 (a) \hspace*{0.45\linewidth} (b)
\caption{{\bf $\nu_\tau$-$\nu_\tau$ interaction in inverted hierarchy:} (a) The shaded region corresponds to the specific flux of the neutrinos allowed by the 1-$\sigma$ 
ranges of all the parameters in table~\ref{tab:inv}. The values of $m_\phi,g_\phi$ and $\gamma$ has been fixed to their bestfit values (in 
table~\ref{tab:inv}) for all the solid red line, orange dashed line and the pink dotted line. However, the solid red line corresponds to 
the bestfit value of $m_0$; the orange dashed line corresponds to the $m_0$ value fixed at zero and the pink dotted line 
corresponds to the lowest 1-$\sigma$ allowed value of $m_0$. (b) The allowed 1-$\sigma$ and 2-$\sigma$ regions in the $m_\phi$-$g_\phi$ 
plane from \ic data are shown along with the cosmological bound (``Planck+BAO+HST'') for $\nu_\tau$-$\nu_\tau$ interaction in the inverted hierarchy. The allowed 
region for \ic has no overlap with the strong self-interaction (SI) band.  
}\label{fig:tau_inv_ic}
 \end{figure}
 
\paragraph{$\nu_\tau$-$\nu_\tau$ interactions:}

 In case of inverted hierarchy, for the $\nu_\tau$-$\nu_\tau$ interaction, number of dips in the \ic energy range depends on the $m_0$ 
value.  For very small value of $m_0$ (close to zero) we get one dip in the \ic energy range. Similarly, in the case of larger values of $m_0$, when all the three mass eigenstates become almost degenerate, we find only one dip in the \ic energy range. On the other hand, for the values of $m_0$, which are not very close to zero and also not very large to make all three mass eigenstates degenerate, we can get two dips in the \ic 
energy range. 
The bestfit values along with 1-$\sigma$ error for all the parameters ($m_\phi,g_\phi,\gamma$ amd $m_0$) are given in table~\ref{tab:inv}. In 
\fig{fig:tau_inv_ic}, the specific flux in \ic and the corresponding allowed parameter space for $m_\phi$-$g_\phi$ have been shown. We have also plotted three different lines, the solid red, orange dashed and dotted pink line, for the specific flux in \ic which show very different features. The orange dashed line is for $m_0=0$ eV and all other parameters are fixed at their best fit 
values. We can see from the 
\fig{fig:tau_inv_ic}-(a) that this line represents the one dip solution. As we increase the $m_0$ value, the orange dashed line moves towards 
the pink dotted line. In case of the pink dotted line, we have kept the value of $m_0$ fixed at lowest 1-$\sigma$ value. It can be clearly seen from the \fig{fig:tau_inv_ic}-(a) that pink dotted 
line can explain two dips in \ic data at two different energies. If 
the neutrino mass is increased further the pink dotted line  moves towards the solid red line which corresponds to the 
bestfit values of the all parameters presented in table~\ref{tab:inv}.    
Therefore, we can clearly see from the \fig{fig:tau_inv_ic}-(a) that the solid red and orange dashed line represent the one dip solutions in the 
\ic energy range, whereas the pink dotted line represents the two dip solution. The shaded region in the 
\fig{fig:tau_inv_ic}-(a) corresponds to the specific flux of the neutrinos for the 
maximum allowed values of all the parameters ($m_\phi,g_\phi,\gamma$ and $m_0$) within the 1-$\sigma$. Once again, It is quite evident from 
\fig{fig:tau_inv_ic}-(b) that the \ic allowed parameter space for 
$m_\phi$-$g_\phi$ is inconsistent with SI bound. However, \ic allowed parameter space for $m_\phi$-$g_\phi$ is consistent with MI bound at 
1-$\sigma$ level. We have also presented the \ic allowed parameter space for $m_\phi$-$m_0$ in \fig{cmb_mass_bound_inverted}-(b). The 
1-$\sigma$ maximum value of $m_0$ allowed by `` Planck+BAO+HST" data is 0.069 eV for moderate $\nu_\tau$-$\nu_\tau$ interaction in case of 
inverted hierarchy. This cosmological disallowed region is shown in the green shade in \fig{cmb_mass_bound_normal}-(b) which implies that a substantial 
part of \ic allowed $m_0$ values is ruled out by the cosmological bound.

\section{Conclusions}\label{Conclusions}
In this paper we have studied the self-interaction between the neutrinos in the context of $H_0$ tension and the observed dips in the 
neutrino flux at \ic. We have shown that the flavour specific interaction and the universal interaction does not affect CMB power spectrum 
very much differently. Even the inverted hierarchy and the normal hierarchy do not have much distinguishable effect. The bound on the self 
interaction parameter $G_{\rm eff}$ from the ``Planck+BAO+HST'' data shows a bimodal feature in its distribution which is consistent with 
the earlier studies. The bestfit values of $G_{\rm eff}$ from the MCMC analysis are also similar for different types of interaction in 
different hierarchies. If self-interaction is not considered in the neutrinos, the allowed value of $N_{\rm eff}$ predicts $H_0$ much lower than the value obtained from HST measurement~\cite{Aghanim:2018eyx}. Whereas, for the cosmological model having self-interacting neutrinos, the $H_0$ value obtained in a joint analysis of Planck CMB, BAO, and HST data is higher than that obtained from the $\Lambda$CDM model. In that case, the obtained $H_0$ overlaps with the $H_0$ from the HST measurement. However, it should be noted that for the self-interacting neutrino model, the $H_0$ inferred from the Planck CMB data alone does not change much from the $H_0$ values obtained within the $\Lambda$CDM model.


The effect of the flavour specific and the universal interaction on the total flux of astrophysical neutrinos in \ic is quite different. 
We have plotted different ways of fitting \ic data with these interactions and constrained the parameter space of self-interaction mediator 
and the neutrino mass. We find that the dips in between 400 TeV -1 PeV and at around $6$ PeV can be simultaneously explained using the neutrino self-interaction in flavour specific cases\footnote{The recent observation~\cite{IceCube:2021rpz} of the Glashow resonance event by the \ic collaboration might impact this conclusion. However, we expect the change in constrained parameter space will not affect the paper's conclusion. This is because of the observed flux at around $6$ PeV in ref.~\cite{IceCube:2021rpz} overlaps significantly with the 1-$\sigma$ band shown in our plots (\fig{fig:uni_norm_ic}-a, \fig{fig:tau_norm_ic}-a, \fig{fig:uni_inv_ic}-a, and \fig{fig:tau_inv_ic}-a)}. However, in case of universal interaction it is impossible to explain two dips simultaneously and by adjusting the parameters, only one dip can be explained. Moreover, by suitably adjusting the lowest neutrino mass, two different dips in flavour specific cases can also be merged into a single dip. We find that there is no distinguishable difference in the features of the neutrino flux line for 
$\nu_e$-$\nu_e$, $\nu_\mu$-$\nu_\mu$ and $\nu_\tau$-$\nu_\tau$ interaction. Therefore, all the analysis and the results presented here for the 
$\nu_\tau$-$\nu_\tau$ interaction are equally applicable to the $\nu_e$-$\nu_e$ and $\nu_\mu$-$\nu_\mu$ interaction cases.  

Hierarchies also play an important role in determining the position and the shape of the absorption dips in the neutrino flux at \ic. In 
the case of universal interaction and normal hierarchy, two very small dips can occur for the lowest neutrino mass being zero. However, 
for the same case in the inverted hierarchy, there can be only one small dip if $m_0$ is fixed at zero. In case of $\nu_\tau$-$\nu_\tau$ 
interaction, normal hierarchy can produce two dips in the two above-mentioned energy values (around 500 TeV and $6$ PeV) for the value of $m_0$ fixed at zero. However, in the case of 
inverted hierarchy, that same feature requires a small but non-zero lowest neutrino mass. 

Since, the value of mediator mass ($m_\phi$) changes the position of the dips and the value of the interaction strength ($g_\phi$) changes 
the sharpness of the dips, the MCMC analysis provide quite strict bound on those values. Larger values of $g_\phi$ beyond the obtained 
bound 
can produce sharper dips in the resonance energies, but it will not be able to explain the absorption feature throughout the specified 
energy range of a certain bin in the \ic data. We found that the cosmological bound on the neutrino self-interaction parameters in 
strong interaction (SI) region, which are inferred from the joint analysis of Planck, HST and BAO data, are inconsistent with the parameter 
space inferred from the \ic data. The preferred parameter space of $g_\phi$-$m_\phi$ by the \ic data can only be slightly consistent with the moderate 
interaction (MI) allowed by the cosmological data set. More specifically, only the $\nu_\tau$-$\nu_\tau$ interaction in the inverted hierarchy 
prefers $g_\phi$-$m_\phi$ parameter space that is in 1-$\sigma$ concordance with cosmological constraints. For all other cases of neutrino 
self-interactions studied in this paper allowed parameter space from \ic data and cosmological data matches at 2-$\sigma$ level only. We 
have also plotted the allowed region of flux line beyond 10 PeV neutrino energy in our results. Therefore, it serves as a prediction for 
upcoming data in \ic experiment in ultra high energies.

\section* {Acknowledgment}
We acknowledge the computation facility, 100TFLOP HPC Cluster, Vikram-100, at Physical Research Laboratory, Ahmedabad, India. The authors sincerely thank the referee for their insightful comments and suggestions that helped in improving the structure and quality of this paper. Some part of this manuscript during the last stage of the review was completed during PP's tenure at Indian Institute of science for which PP acknowledge IOE-IISc fellowship program for financial assistance.

\section*{Appendix}
\appendix
\section{Numerical details}\label{app:numerical}
The MCMC analysis performed in the paper for the \ic data is based upon the Metropolis-Hastings algorithm~\cite{Hastings:1970aa}. The 
likelihood function 
used in the analysis assumes a generalized likelihood function for asymmetric error, following ref.~\cite{Barlow:2004wg} which can be 
written as
\begin{eqnarray}
 \ln[L] = \sum-{1\over 2}\left(\hat{x}-x\over \sigma+\sigma'(\hat{x}-x)\right)^2\, ,
\end{eqnarray}
where, 
\begin{eqnarray}
 \sigma = {2\sigma_+ \sigma_{-}\over \sigma_+ + \sigma_{-} } {\, \,\rm and \,\,} \sigma' = {\sigma_+ - \sigma_{-}\over \sigma_+ + 
\sigma_{-}}\, ,
\end{eqnarray}
with $\sigma_+$ and $\sigma_{-}$ are the positive and negative errors.
In case of symmetric errors $\ln[L]$ turns out to be $-\chi^2$. 
In the likelihood function we used the binned flux data of ref.~\cite{Kopper:2017zzm} as observational values. Theoretical values are 
calculated from the solution of \eqn{eqn:phi} for 63 different points in the \ic energy range where we have taken 9 points from each bin in 
equally spaced log space of $E_0$.  

We have used the Gaussian priors for all the parameters and the corresponding values are given 
table~\ref{tab:prior_ic}. We have also set the maximum and minimum values at 7.4 and 6.2 respectively for 
$\log_{10}({m_\phi/{\rm eV}})$. Lowest neutrino mass $m_0$ has also been assigned with maximum and minimum values of $0.15$ eV and $0$ 
respectively.
\begin{table}[b]
\begin{center}
\begin{tabular} { | l | c | c |}
\hline
Parameter & mean & 1-$\sigma$\\
\hline
{\boldmath$m_0   ({\rm eV})       $} & 0.05 & 0.02 \\

{\boldmath$\log_{10}({m_\phi/{\rm eV}})       $} & 6.7 & 0.05 \\

{\boldmath$\log_{10}{g_\phi}      $} & -1.5 & 0.05 \\

{\boldmath$\gamma         $} & 2.35 & 0.05 \\
\hline
\end{tabular}
\end{center}
\caption{Priors used in MCMC with \ic data}\label{tab:prior_ic}
\end{table} 

 The cosmological parameters of interest in this paper are the six standard cosmological parameters, effective number of relativistic 
degrees of freedom $N_{\rm eff}$, lowest neutrino mass $m_0$ and effective coupling constant of self-interaction ($G_{\rm eff}$). The six 
standard cosmological parameters are: the
fraction density of cold dark matter and baryonic matter at present multiplied by square of the reduced Hubble parameter ($\omega_{\rm cdm}$ 
and $\omega_b$ respectively), acoustic scale of baryon acoustic oscillation ($\theta_s$), amplitude and the spectral index of the primordial 
density perturbations ( $A_s$ and $n_s$ respectively) and optical depth to the epoch of re-ionization ($\tau_{\rm reion}$). 
These nine parameters have been varied in this analysis and the corresponding priors for these parameters are given in table~\ref{tab:prior}.
\begin{table}[h]
\begin{center}
\begin{tabular} { | l | c | c |}
\hline
Parameter & mean & 1-$\sigma$\\
\hline
$\omega_b$ & $2.2377\times 10^{-2}$ &  $0.015\times 10^{-2}$\\
$\omega_{\rm cdm}$  & 0.12010 & 0.0013 \\
$100\theta_s$  & 1.04110  & 3e-4 \\
$\ln (10^{10}A_s)$ & 3.0447 & 0.015 \\
$n_s$          & 0.9659 & 0.0042 \\
$\tau_{\rm reio}$   & 0.0543 &   0.008\\ 
$\log_{10}({G_{\rm eff}})$            & -1.5 & 0.2\\
$m_0$ & 0.008 & 0.005 \\
$N_{\rm eff}$ & 3.75 & 0.1 \\
\hline
\end{tabular}
\end{center}
\caption{Priors used in MCMC with Planck, BAO and HST data.}\label{tab:prior}
\end{table}
 We have used Gaussian prior for our purpose. In case 
of $\log_{10}({G_{\rm eff}})$, we have assigned the maximum and minimum values -5.0 and -0.1 respectively. Lowest neutrino mass $m_0$ has been varied in the range [0, 0.2] eV and $N_{\rm eff}$ has been varied in the range [3,5]. We have also assigned a minimum value of $\tau_{\rm reio}$ at $0.04$. While sampling the parameter space, we increased temperature of the chains by a factor of three following ref~\cite{Oldengott:2017fhy} for proper sampling. 

To show the comparison between different models under consideration in this work, we report the maximum value of likelihood (minimum of $-\log$(likelihood)) in Table~\ref{tab:lik_values}. It is evident from this table that there is a significant improvement in the -$\log$(likelihood) values after the inclusion of self-interaction in neutrinos for the joint analysis of Planck, BAO, and HST data.

\begin{table}[h]
\begin{center}
\begin{tabular} { | c | c |}
\hline
Model & $-\log$(likelihood)\\
\hline
$\Lambda$CDM  & 794.84   \\
Universal Interaction (NH) & 263.89  \\

Universal Interaction (IH) & 263.51 \\

$\nu_\tau$-$\nu_\tau$ 
Interaction (NH)  & 264.02  \\

$\nu_\tau$-$\nu_\tau$ 
Interaction (IH)  & 236.58  \\
\hline
\end{tabular}
\end{center}
%
\caption{ The $-\log$(likelihood) values obtained from the MCMC analysis of the cosmological model with universal and 
$\nu_\tau$-$\nu_\tau$ self interactions and the standard $\Lambda$CDM model.}\label{tab:lik_values}
\end{table}

\bibliographystyle{JHEP}
\bibliography{sinu_icecubeNotes.bib}
\end{document}